\documentclass[superscriptaddress,showpacs,showkeys, onecolumn, notitlepage]{revtex4-1}

\usepackage{amsmath, amsthm, amssymb}

\usepackage{upgreek}
\usepackage{graphicx}

\usepackage{hyperref}
\hypersetup{
    colorlinks=true,
    citecolor=blue,
    linkcolor=black,
    filecolor=blue,      
    urlcolor=blue,
}

\usepackage{natbib}

\usepackage{ulem}
\usepackage{bm}
\usepackage{color}
\usepackage{times}

\usepackage{amsmath}
\usepackage{amsfonts}
\usepackage{amssymb}

\makeatletter
\renewcommand\thesection{}
\renewcommand\thesubsection{\@arabic\c@section.\@arabic\c@subsection}
\makeatother

\expandafter\ifx\csname package@font\endcsname\relax\else
 \expandafter\expandafter
 \expandafter\usepackage
 \expandafter\expandafter
 \expandafter{\csname package@font\endcsname}%
\fi
\begin{document}

\renewcommand{\figurename}{}
\renewcommand{\thefigure}{Figure~\arabic{figure}}
\setcounter{figure}{0}

\renewcommand{\theequation}{\arabic{equation}}
\setcounter{equation}{0}

\renewcommand{\tablename}{}
\renewcommand{\thetable}{Table~\arabic{table}}
\setcounter{table}{0}

\renewcommand{\thesection}{\arabic{section}}
\setcounter{section}{0}

\title{Exotic Multifractal Conductance Fluctuations in Graphene }

\author{Kazi Rafsanjani Amin}
\affiliation{Department of Physics, Indian Institute of Science, Bangalore, Karnataka, India 560012}
\author{Samriddhi Sankar Ray}
\affiliation{International Centre for Theoretical Studies, Tata Institute of Fundamental Research, Bangalore, Karnataka, India 560089}
\author{Nairita Pal}
\affiliation{Department of Physics, Indian Institute of Science, Bangalore, Karnataka, India 560012}
\author{Rahul Pandit}
\affiliation{Department of Physics, Indian Institute of Science, Bangalore, Karnataka, India 560012}
\author{Aveek Bid}
\email{aveek@iisc.ac.in}
\affiliation{Department of Physics, Indian Institute of Science, Bangalore, Karnataka, India 560012}

\begin{abstract}
In quantum systems, signatures of multifractality are rare.  They have  been found only in the multiscaling of eigenfunctions at critical points. Here we demonstrate   multifractality in the magnetic-field-induced universal conductance fluctuations of the conductance in a quantum condensed-matter system, namely, high-mobility single-layer graphene field-effect transistors.  This multifractality decreases as  the temperature increases or as doping moves the system away from the Dirac point. Our measurements and analysis  present evidence for an incipient Anderson-localization near the Dirac point as the most plausible cause for this multifractality.  Our experiments suggest that multifractality in the scaling behaviour of local eigenfunctions are reflected in macroscopic transport coefficients.  We  conjecture that an incipient Anderson-localization transition may be the origin of this multifractality. It is possible that multifractality is ubiquitous in transport properties of low-dimensional systems. Indeed, our work suggests that we should look for multifractality in transport in other low-dimensional quantum condensed-matter systems. 
\end{abstract}

\maketitle

\section{Introduction}

One of the unsolved problems in  single-layer graphene (SLG) is the nature of the electronic wave-function near the charge-neutrality (Dirac) point. In principle, the charge-carrier density of  SLG  should be continuously tunable, down to zero, leading  to the largely unexplored regime of extremely weak interactions in a low-carrier-density system.  The interaction parameter $r_{\mathrm{s}}$, which parametrizes the ratio of the average inter-electron Coulomb interaction energy to the Fermi energy, turns out to be independent of charge-carrier density in the case of SLG where $r_{\mathrm{s}} = e^2/(\kappa\hbar v_{\mathrm{F}})$. Here, $\kappa$ is the dielectric constant of the surrounding medium and $v_{\mathrm{F}}$ is the Fermi velocity. In the case of SLG on an SiO$_2$ substrate, $r_{\mathrm{s}}$ = 0.8. Thus SLG is a very weakly interacting system, when compared to other conventional two-dimensional systems such as GaAs/AlGaAs and Si inversion layers, where the values of $r_{\mathrm{s}}$ are typically much higher. 
A na\"ive application of the scaling theory to such a system in this regime would predict Anderson localization--- a disorder-driven quantum phase transition, leading to a complete localization of the charge-carriers, and thence, to an  insulator~\cite{RevModPhys.80.1355, PhysRevB.50.7526, PhysRevB.74.041403, PhysRevLett.97.236801, Katsnelson2006}.  Indeed, theoretical calculations for graphene indicate that intervalley scattering can lead to changes in the local and averaged electronic density of states (DOS) with the creation of localized states~\cite{PhysRevLett.96.036801, PhysRevLett.116.126804}. For example, graphene-terminated SiC (0001) surfaces undergo an Anderson-localization transition  upon dosing  with small amounts of atomic hydrogen \cite{PhysRevLett.103.056404}. This  intervalley-scattering-induced localization is most  effective near the Dirac point, where the screening of the impurity scatterers is negligible \cite{PhysRevLett.116.126804}.  Experiments,  however, find the appearance of a minimum conductance value at the Dirac point with $\sigma_{\textrm{min}}\approx 2e^2/\pi\hbar$, instead of a diverging resistance which is the hallmark of a truly localized state; notable exceptions being carrier localization~\cite{Jung2011} and an Anderson-localization transition in bi-layer graphene heterostructures~\cite{Ponomarenko2011}.

By studying the scaling behaviour of the universal conductance fluctuations (UCF), we look for signs of charge-carrier localization near the Dirac point in ultra-high mobility SLG and uncover, as a result, an exotic multifractal behaviour in the UCF. Multifractality, characterized by an infinite number of scaling exponents, is ubiquitous in classical systems.  Since the pioneering work of Mandelbrot~\cite{mandelbrot19827he}, the detection and analysis of multifractal scaling in such systems have enhanced our understanding of several complex phenomena, e.g.,  the dynamics of the human heart-beat~\cite{Ivanov1999}, the form of critical wave-functions at the Anderson-localization transition~\cite{RevModPhys.80.1355}, the time series of the Sun's magnetic field~\cite{lawrence1993multifractal}, in medical-signal analysis (for instance, in pattern recognition, texture analysis and segmentation)~\cite{6602370}, fully-developed turbulence and in a variety of chaotic systems~\cite{frisch1995turbulence,0305-4470-17-18-021}.  In condensed-matter systems, signatures of multifractality are usually sought in the scaling of eigenfunctions   at critical points~\cite{0305-4470-19-8-004, JANSSEN19981, Mirlin2000259,   PhysRevB.51.663, PhysRevB.91.085427, PhysRevLett.111.066601,   PhysRevB.87.125144,2053-1583-1-1-011009}. Despite compelling theoretical predictions~\cite{Facchini2007266, PhysRevB.79.205120,PhysRevLett.54.1718, PhysRevB.43.3601,PhysRevLett.72.3582,PhysRevLett.81.4696}, there are no reports of the successful observation of multifractality in transport coefficients.

Simple fractal conductance fluctuations, on the other hand, have been observed in several condensed-matter systems \cite{PhysRevLett.77.3885, PhysRevLett.87.036802, PhysRevLett.80.1948,  PhysRevB.54.10841, PhysRevLett.77.3885, PhysRevLett.87.036802, 1742-6596-109-1-012035}.  These arise through semi-classical electron-wave interference processes whenever a system has mixed chaotic-regular dynamics~\cite{MEISS1986387, PhysRevLett.59.2503, PhysRevB.54.10841, PhysRevLett.77.3885}. In all these systems, a primary prerequisite for the observation of fractal transport is that the electron dynamics be amenable to semi-classical analysis - the fractal nature of conductance fluctuations seen in these systems disappears as the system is
driven deep into the quantum limit.  
For such a mixed-phase semi-classical system, the graph of conductance ($G$)  versus  an externally applied magnetic field ($B$) has the same statistical properties as a Gaussian random process with increments of mean zero and variance $(\Delta B)^\gamma$. These   processes are known as fractional Brownian motion and have the property that their graph is a fractal of dimension  $D_{\mathrm{F}} = 2 - \frac{\gamma}{2}$~~\cite{PhysRevLett.77.3885, PhysRevB.54.10841},  with $1\leq D_{\mathrm{F}}\leq 2$.

A semi-classical description, however, breaks down in the case of materials where the charge-carriers obey a Dirac dispersion relation, e.g., in graphene and topological insulators. In this report, we ask:
can we find signatures of multifractality  in a quantum system through transport measurements, specifically through the statistical properties of the graph of conductance fluctuations versus an external parameter like the magnetic field? 
We address this question by studying in detail the statistics of conductance fluctuations in high-mobility SLG-FET devices as a function of the perpendicular magnetic field over a wide range of temperature and doping levels. We report, for the first time, the  occurrence of multifractality in the UCF in  high mobility graphene devices deep in the quantum limit. Our measurements and analysis  suggest at an incipient Anderson localization transition in graphene near the Dirac point.

\section{Results}

\begin {figure*}[!t]
  \begin{center}
    \includegraphics[width=.7\textwidth]{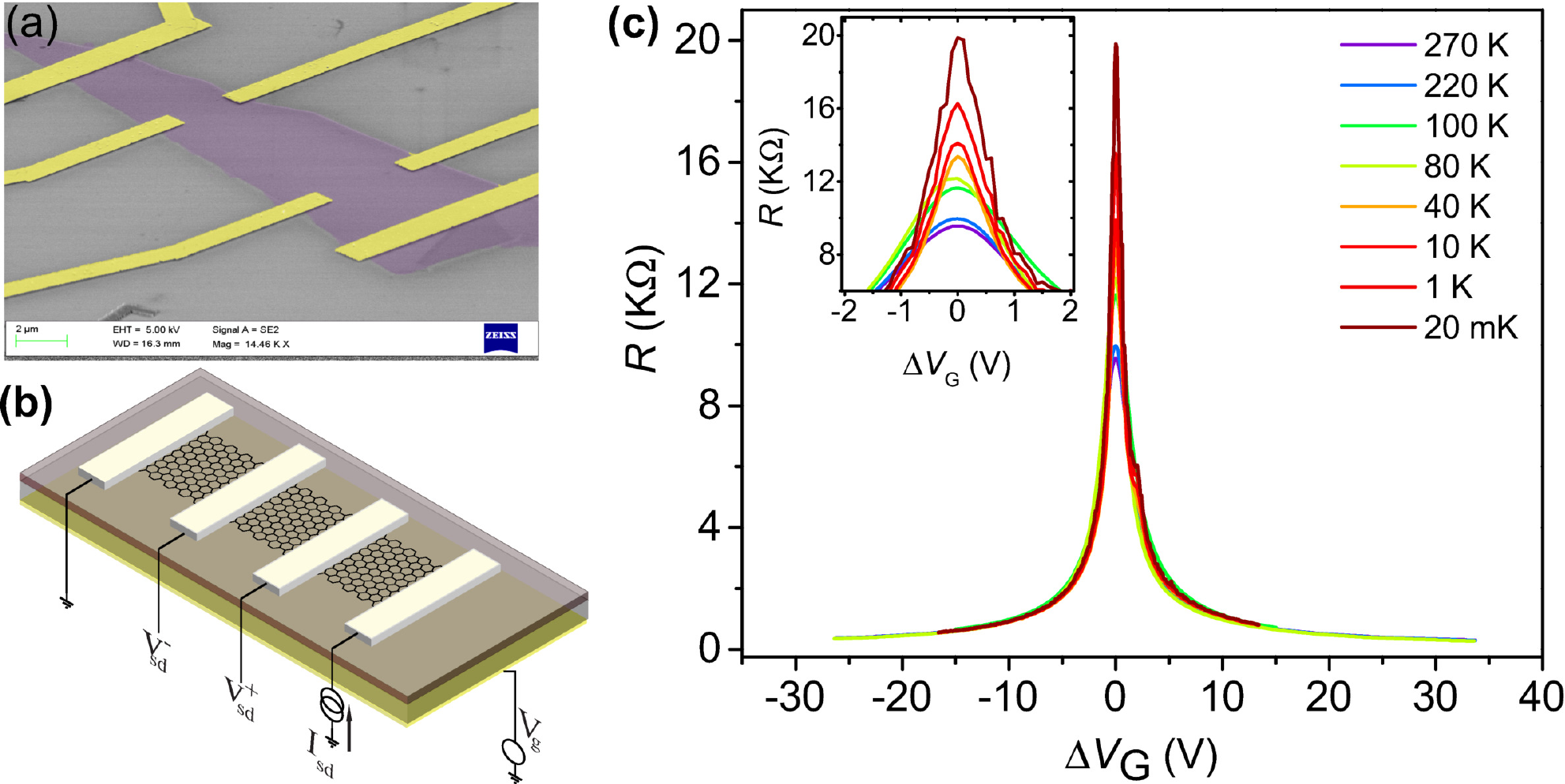} 
    \small { \caption{\textbf{Device structure and characteristics.} (a) False-colour SEM image and (b)  schematic device configuration of our SLG-FET. The scale-bar in (a) is 2~$\mu m$.  
        (c) Plot of $R$ versus $\Delta V_{\mathrm{G}}$, measured at different temperatures. The values of the temperatures at which the data were collected are mentioned in the legend. In the inset, we plot the same data, zoomed in near the Dirac point ($|\Delta V_{\mathrm{G}}| \leq 2$~V).  
        \label{fig:device} }}
  \end{center}
\end {figure*}

\subsection*{Measurement of Universal Conductance Fluctuations.}

\begin {figure*}[!t]
  \begin{center}
    \includegraphics[width=\textwidth]{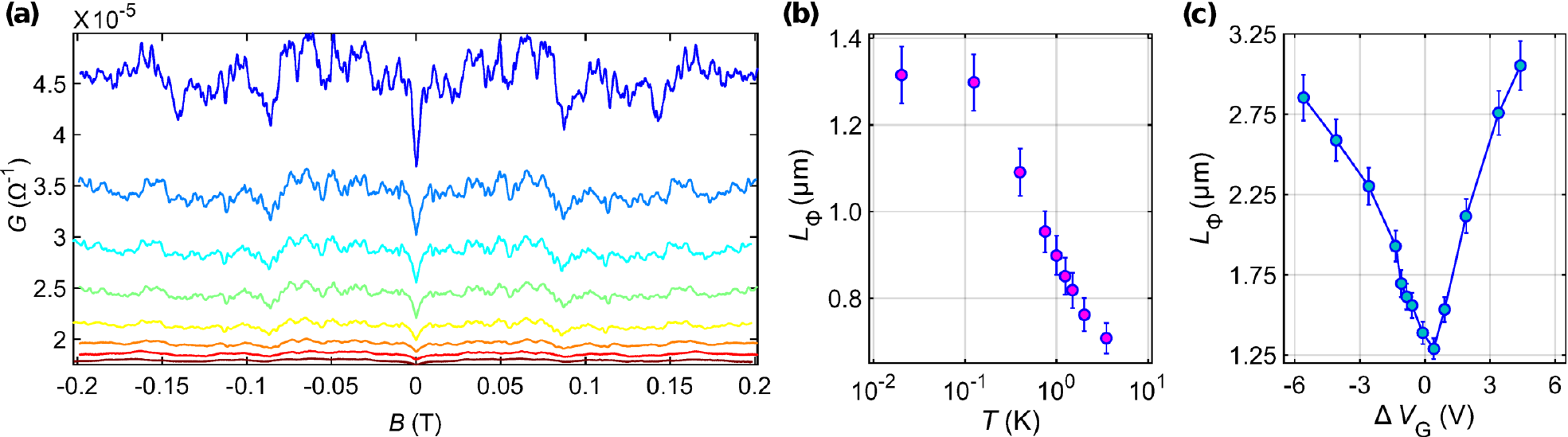} 
    \small { \caption{\textbf{ UCF in graphene.} (a)  Illustrative 	plots of the magnetoconductance $G$ versus the magnetic field $B$ measured at 	0.02~K, 0.05~K, 0.07~5K, 0.10~K, 0.20~K, 0.50~K, 1.00~K, and 2.00~K (from top to bottom - curves for different temperatures have been shifted vertically for clarity). The data shown here are from the device G28M6 measured close to the Dirac point ( $\Delta V_{\mathrm{G}}=0.2$~V). 
        (b)  Plot of the phase-coherence length $L_\upphi$ versus the temperature $T$ extracted from the UCF measured at $\Delta V_{\mathrm{G}}=0.2$~V. (c) Plot of $L_\upphi$ versus  $\Delta V_{\mathrm{G}}$, extracted from the  UCF measured at $T=$ 20 mK. The error-bars in (b) and (c) represent the standard deviation in $L_\upphi$.
        \label{fig:ucf_T} }}
  \end{center}
\end {figure*}

SLG-FET devices with mobilities in the range $[20,000 - 30,000]$ cm$^2$V$^{-1}$s$^{-1}$ were fabricated on SiO$_2$ substrates by
mechanical exfoliation from natural graphite, followed by conventional, electron-beam lithography~\cite{novoselov2005two}[\ref{fig:device}~(a-b)]. 
We begin with the dependence of the resistance ($R$) on the gate voltage ($V_{\mathrm{G}}$),  for the device G28M6. In \ref{fig:device}(c) we show plots of  $R$ versus $\Delta V_{\mathrm{G}} = V_{\mathrm{G}} -V_{\mathrm{D}}$, where $V_{\mathrm{D}}$ is the Dirac point,  measured at different temperatures ($T$). The  high mobility and the position of the charge-neutrality point very close to $V_{\mathrm{G}}=0$~V attests to the high quality of the devices.  We measured the magnetoconductance ($G$) as a function of the magnetic field $(\textbf{B}=(0,0,B) ) $, applied perpendicular to the plane of the device , in the range $-0.2$~T $\le$ $B$ $\le$ $0.2$~T. The presence of UCF was confirmed by the appearance of reproducible, non-periodic, but magnetic-field-symmetric, oscillations in $G$.  The measurements were performed on multiple devices, over a wide range of $V_{\mathrm{G}}$ and $T$ .  We find our UCF data to be in excellent agreement with previous studies of magnetoresistance and conductance fluctuations in single-layer graphene~\cite{0953-8984-22-20-205301, Berger1191, Bohra2012, PhysRevLett.100.056802, Horsell20091041}. \ref{fig:ucf_T}(a) shows illustrative plots of $G(B)$ with the Fermi energy ($E_{\mathrm{F}}$)  maintained very close to the Dirac point  [$\Delta V_{\mathrm{G}}  ~\simeq 0$] for the device G28M6. The  data for other devices are similar and are shown in the  Supplementary Figure~1 [see Supplementary Note~1]. 
At low values of $|B|$, near the minimum of $G$ at $B=0$, weak-localization corrections are visible [\ref{fig:ucf_T}(a)].  As we move away from $B=0$, the conductance fluctuations become prominent.  The amplitudes of the UCF peaks, and the values of the charge-carrier phase-coherence length $L_\upphi$, obtained from the variance of the UCF [\ref{fig:ucf_T}(b)], decrease with increasing temperature because of  thermal dephasing [see Supplementary Note~2].   
 The temperature dependence of the intervalley-scattering length and the intravalley-scattering length, extracted from weak-localization measurements at $T=$20 mK and $\Delta V_{\mathrm{G}}=$0.2 V, are shown in the  Supplementary Figure~4 [see Supplementary Note~3].  We find them to be in excellent agreement with previous studies of localization in single-layer graphene~\cite{PhysRevLett.98.176805,PhysRevLett.100.056802}. 
We also observe an increase in $L_\upphi$ with increasing $|\Delta V_{\mathrm{G}}|=|V_{\mathrm{G}}-V_{\mathrm{D}}|$~[\ref{fig:ucf_T}(c)], which we attribute to the increase in the screening of impurities  by charge carriers~\cite{0953-8984-22-20-205301}. 
We note an apparant saturation of  $L_\upphi$ below a temperature of $\simeq100$~mK. The saturation of the phase-coherence length ($L_\upphi$) with decreasing temperature is an issue that has been at the forefront of research in several other semiconducting materials including doped Si~\cite{PhysRevLett.91.066604,PhysRevB.55.R13452,PhysRevB.68.085413}. There are many effects, e.g. the presence of magnetic impurities~\cite{PhysRevB.68.085413} or finite size, which can lead to a saturation of $L_\upphi$. We  note here that a decoupling of the electron and lattice temperature leading to a saturation of the $L_\upphi$ is also possible. However, the data shown in \ref{fig:device}(c) and \ref{fig:ucf_T}(a) show a continuous evolution of both the conductance and conductance fluctuations down to 20 mK. This rules out any saturation of the electron temperature down to 20 mK, and hence, the observed saturation of $L_\upphi$ below a certian temperature is not an experimental artifact [see Supplementary Note~3].
In a separate set of measurements, we obtain the magnitude of the UCF, at a given magnetic field,  by sweeping over $V_{\mathrm{G}}$ and calculating the rms value of the fluctuations.  We  found that this quantity   decreased sharply with increasing magnetic field [see Supplementary Note~4, Supplementary Figure~5] in conformity with theoretical predictions~\cite{Liu2016}.

\begin {figure*}[!t]
  \begin{center}
    \includegraphics[width=.75\textwidth]{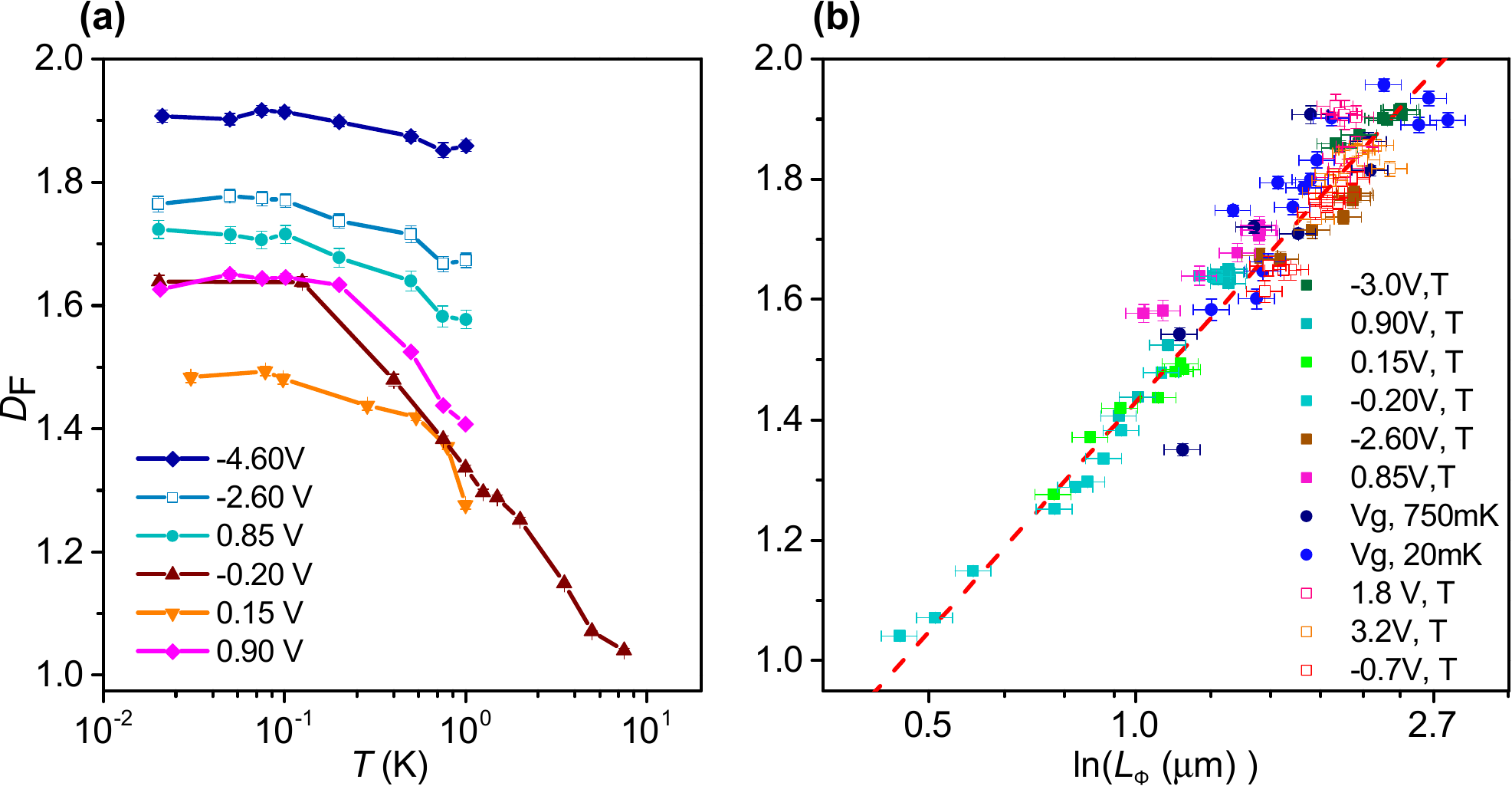}
    \small {\caption{\textbf{Fractal dimension of UCF in graphene.}  (a) Plots of the fractal dimension $D_{\mathrm{F}}$ versus the temperature $T$ for different values of  $\Delta V_{\mathrm{G}}$. The error-bars represent the standard deviation in $D_{\rm F}$.
        (b) $D_{\mathrm{F}}$ versus the charge-carrier phase-coherence length $L_\upphi$ on a semi-log scale; the data points, with error bars representing standard deviation in $D_{\rm F}$ and $L_\upphi$., are from two different devices (filled symbols: G28M6; open symbols: G30M4) and for the  following parameter ranges: $20~\rm{mK} <$ $T$ $< 10~\rm{K}$ and $-4.5~\rm{V}< \Delta V_{\mathrm{G}} < 4.5$~V. The numbers in the legend refer to the ($V_{\mathrm{G}},T$) parameter space. The red  curve is $D_{\mathrm{F}} \propto \ln (L_\upphi)$ (see text).
        \label{fig:lphi_df} }}
  \end{center}
\end {figure*}

\subsection*{Analysis of Fractal scaling of the UCF.}

UCF represents quantum correction to Drude conductivity arising from the interference of electronic wavefunctions; and it is  fingerprint of the disorder configuration in the conducting channel. Besides information about the phase-coherence of the charge carriers, it can also provide crucial insights into the electron dynamics and distribution of eigenstates through a scaling-dimension analysis of the magnetoconductance traces~\cite{PhysRevB.54.10841, PhysRevLett.77.3885,  PhysRevLett.87.036802}.   We first compute the simple fractal dimensions $D_{\mathrm{F}}$ of the UCF curves via  the Ketzmeric-variance method [Supplementary Note~5]. \ref{fig:lphi_df}(a) shows plots of $D_{\mathrm{F}}$ versus $T$  for the device G28M6. At very low $T$ and small $|\Delta V_{\mathrm{G}}|$, we find $1<D_{\mathrm{F}}<2$. With increasing temperature, $D_{\mathrm{F}}$ $\to$1 monotonically. In this high-$T$ regime, the thermal-diffusion length $L_{\mathrm{T}} = \sqrt{\hbar D/k_{\mathrm{B}}T} \ll L_\upphi$, with $D$ the thermal-diffusion coefficient  of the charge carriers; therefore, quasiparticle phase-decoherence, induced by  inelastic thermal scattering, suppresses quantum interference. For large $|\Delta V_{\mathrm{G}}|$, the magnitude of the UCF is comparable to, or smaller than the background electrical noise, so $D_{\mathrm{F}}$$\to$$2$, the value for Gaussian white noise. In \ref{fig:lphi_df}(b), we plot $D_{\mathrm{F}}$ versus $L_\upphi$  for two different devices: G28M6 and G30M4; remarkably, all the data points from these two devices cluster in the vicinity of a curve, with $D_{\mathrm{F}} \propto \ln (L_\upphi)$.  In the limits $L_\upphi \ll L$, the UCF is non-fractal ($D_{\mathrm{F}}\simeq 1$), where as for large $L_\upphi$, the UCF is a fractal, so $1 < D_{\mathrm{F}} < 2$.

\begin {figure*}[!t]
  \begin{center}
    \includegraphics[width=.8\textwidth]{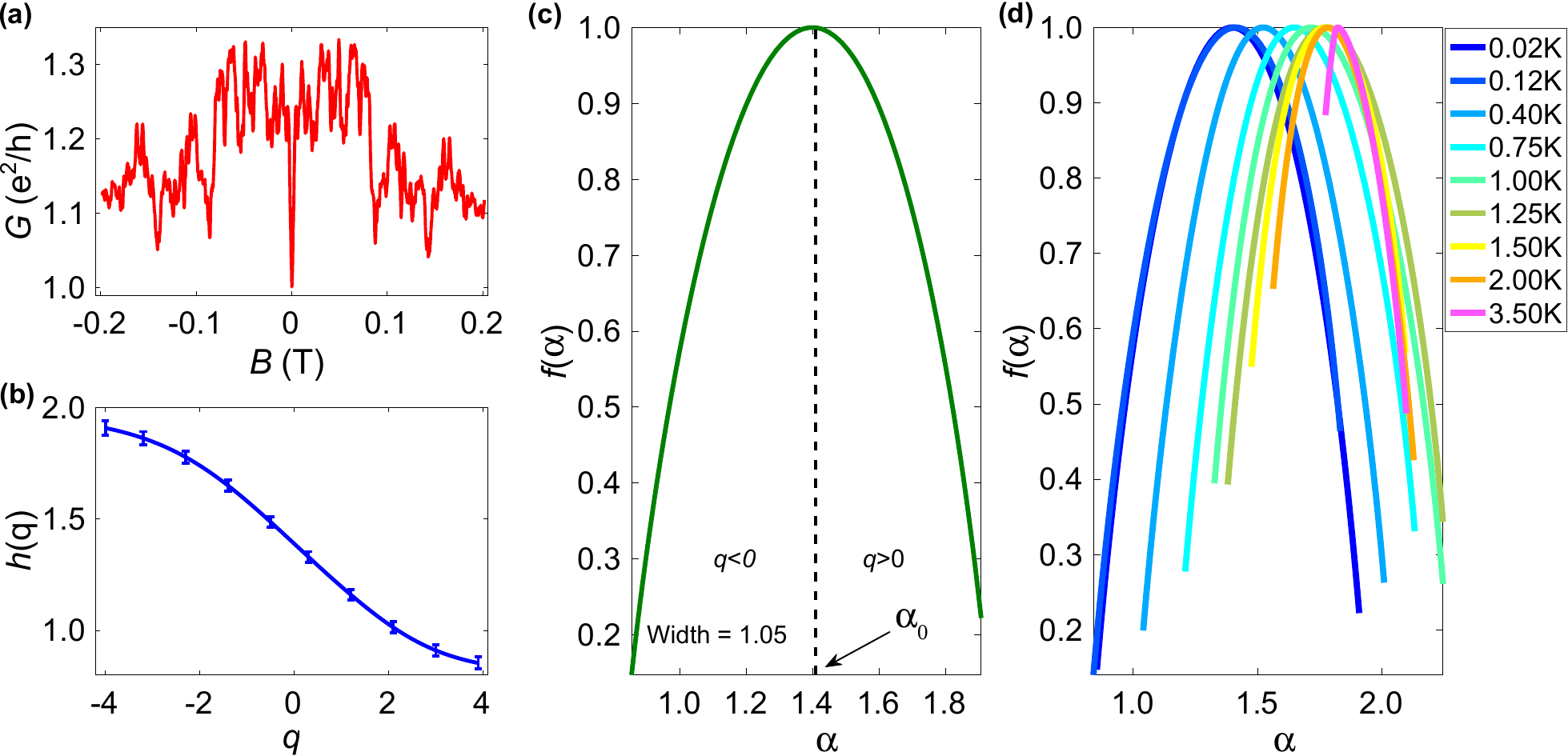}
    \small { \caption{  \textbf{Multifractality of UCF in graphene.} (a) A typical plot of  the magnetoconductance at $T=20$mK,  $\Delta V_{\mathrm{G}} = 0.2$~V for our device G28M6. 
        (b) Plot of the generalized Hurst exponent $h(q)$ versus  $q$, calculated from the UCF data in (a). A few representative error bars, defining the standard deviation in the data, are shown in the plot. 
        (c) Plot of the multifractal singularity spectrum $f(\alpha)$ versus $\alpha$ that follows from the $h(q)$ versus $q$ data plotted in (b). The dotted vertical line marks the location of the maximum of $f(\alpha)$. 
        (d) Plots illustrating the temperature dependence of  $f(\alpha)$ at $\Delta V_{\mathrm{G}} = 0.2$~V for the device G28M6.\label{fig:mf} }}
  \end{center}
\end {figure*}

\subsection*{Analysis of Multifractal scaling of the UCF.}

We build upon the predictions of multifractal scaling of conductance fluctuations in  quantum systems~\cite{Facchini2007266} by carrying out a multifractal detrended fluctuation analysis  [Supplementary Note~6] of our UCF.  The multifractality can be represented in the following two ways: (1) By the generalized Hurst exponent $h(q)$, defined using the order-$q$  moment of the UCF as $\langle {\rm rms}[\Delta G(\Delta B)]^q \rangle^{1/q} \sim [\Delta B]^{h(q)}$, and (2) by the multifractal spectrum $f(\alpha)$, obtained from the Legendre transform  of $h(q)$. For a monofractal function, $h(q)$ has a single, $q$-independent value. For each one of our UCF plots,  we obtain $h(q)$ in the range $-4 \le q \le 4$.  In \ref{fig:mf}(b), we give an illustrative plot of $h(q)$ versus $q$ that we obtain from our magnetoconductance data at 20 mK [\ref{fig:mf}(a)]; $h(q)$ goes smoothly from $\simeq 1.9$, at $q=-4$, to $\simeq 0.85$, at $q=4$; the corresponding $f(\alpha)$ spectrum is plotted in \ref{fig:mf}(c). The singularity spectra have a definite maximum value of 1 (which is the dimension of the support graph). The width of the multifractal spectrum  is defined as $\Delta \alpha  \equiv $ $h(q)_{\rm max}$ -$h(q)_{\rm min}$. The wide range of $h(q)$, or, equivalently, the wide  spectrum ($\Delta \alpha$=1.05) quantifies the multifractality of the UCF. This is the first observation of multifractality of a conductance in any quantum-condensed-matter system and is the central result of our work.

We note that there are two distinct properties of UCF  that can give rise to its multifractal behaviour~\cite{PhysRevE.82.011136,Zhou2009}: (i) a fat-tailed, non-Gaussian distribution of the UCF differences (as a function of $\delta B$)  or (ii) long-range, in the magnetic field $B$, correlations of the fluctuations of $\delta g(B)$.  We have verified that the distribution of our measured $\delta g(B)$  is not log-normal. Having ruled out (i), we now give a convenient test for (ii): we check for multifractality in a data set obtained from a random shuffling [see Supplementary Note~7] of the original sequence of $\delta g(B)$. If the long-range correlations (ii) exist, then signatures of multifractality must be absent in the reshuffled data. Indeed, in our experimental data, we observe a near-complete suppression of multifractality in the shuffled data set with   $h_{\rm shuf}(q) \simeq 0.5$ for all values of $q$ and $\Delta \alpha = $0.05 [Supplmentary Figure~9]. Hence it is reasonable to infer that the multifractality in our UCF can be traced back to long-ranged correlations, that are otherwise difficult to measure.

\ref{fig:mf}(d) shows plots of $f(\alpha)$ for different values of  $T$ and $\Delta V_G=0.2$V. The symmetry of  $f(\alpha)$  about $\alpha_0$ [\ref{fig:mf}(c)] reflects the distribution of fluctuations, about the mean of the UCF. The large-fluctuation (small-fluctuation) segments contribute predominantly to the $q>0$ ($q<0$) part of $h(q)$. The $q>0$ ($q<0$) part of $h(q)$  maps onto the  $\alpha < \alpha_0$ ( $\alpha > \alpha_0$) region of $f(\alpha)$, which is more-or-less symmetric about $\alpha_0$ at low $T$ [\ref{fig:mf}(d)].  As we increase $T$, this symmetry is lost. The magnitude of the skewness $\langle[\delta\alpha]^3\rangle/\langle[\delta\alpha]^2\rangle^{3/2}$, where $\delta\alpha = (\alpha - \alpha_0)$, increases with $T$.  Therefore, as $T$ increases, large-amplitude conductance fluctuations become rarer than small-amplitude fluctuations. This is consistent with our plots of the UCF [\ref{fig:ucf_T}(a)].


\section{Discussion}

A na\"ive characterization of the fractal property of a curve, say by the measurement of one fractal dimension, does not rule out multifractality of this curve, which requires the calculation of an infinite number of dimensions~\cite{PhysRevLett.50.346} (related to $h(q)$). One dimension suffices for monofractal scaling (as, e.g., in the scaling of velocity structure functions in the inverse-cascade region of forced, two-dimensional fluid turbulence~\cite{Kellay2002, Boffetta2012}). Our measurements of the UCF in single-layer graphene show that it is multifractal only if (i) the temperature is low and (ii) $\Delta V_{\mathrm{G}} \simeq 0$. If either one of these conditions is not met, the plot of $\delta g$ versus $B$ is a monofractal [see Supplementary Note~8];  at sufficiently large $T$, $D_{\mathrm{F}} \to$1 and the plots are non-fractal.

What can be the possible origin of our multifractal UCF?  We list three potential causes:
(1) scarring of wave functions
(e.g., because of classically-chaotic billiards~\cite{PhysRevLett.53.1515});  (2) quasi-periodicity in the Hamiltonian induced by a magnetic field~\cite{0370-1298-68-10-305, PhysRevB.14.2239,PhysRevLett.50.1873} and its analogue for graphene~\cite{0953-8984-22-46-465305}; (3) Anderson-localization-induced multifractality~\cite{RevModPhys.80.1355}.  We critically examine each one of these possibilities below and conclude that our results are most compatible with the last of these mechanisms.

While describing a quantum system, whose classical analogue is chaotic,  one encounters scarred wavefunctions, whose intensity is enhanced along unstable,  periodic orbits of the  classical  system. This non-uniform distribution of intensity of wavefunctions results from quasiparticle interference. Quantum scars can lead to pointer states~\cite{PhysRevD.24.1516} with long trapping times, and, consequently, to large conductance fluctuations.  Relativistic quantum scars have been predicted
theoretically in geometrically confined graphene  stadia, which exhibit classical chaos~\cite{PhysRevLett.103.054101}.  However, our devices are not shaped like billiards that are classically chaotic; and the charge carriers are not in the ballistic regime. Therefore, quantum scars cannot be the underlying cause for our multifractal UCF.

Fractal energy spectra can arise in tight-binding problems with an external magnetic field, which can be mapped onto Schr\"odinger problems with quasiperiodic potentials~\cite{0370-1298-68-10-305,PhysRevB.14.2239,PhysRevLett.50.1873,PhysRevB.29.1394}.
It has been argued~\cite{0953-8984-22-46-465305}, therefore, that a fractal conductance can also arise, via Hofstadter-butterfly-type spectra,  in Dirac systems at sufficiently high magnetic fields  $B_{\mathrm{H}} \simeq \phi_0/A_0$ ($\simeq 10^5$ T for our samples), where $\phi_0 = 2 \pi\hbar/e$ is the flux quantum and $A_0$ is the unit-cell area.  Our measurements use very-low-magnitude magnetic fields ($\lesssim 0.2$ T); this rules out a quasi-periodicity-induced multifractal UCF.

\begin {figure*}[!t]
  \begin{center}
    \includegraphics[width=.65\textwidth]{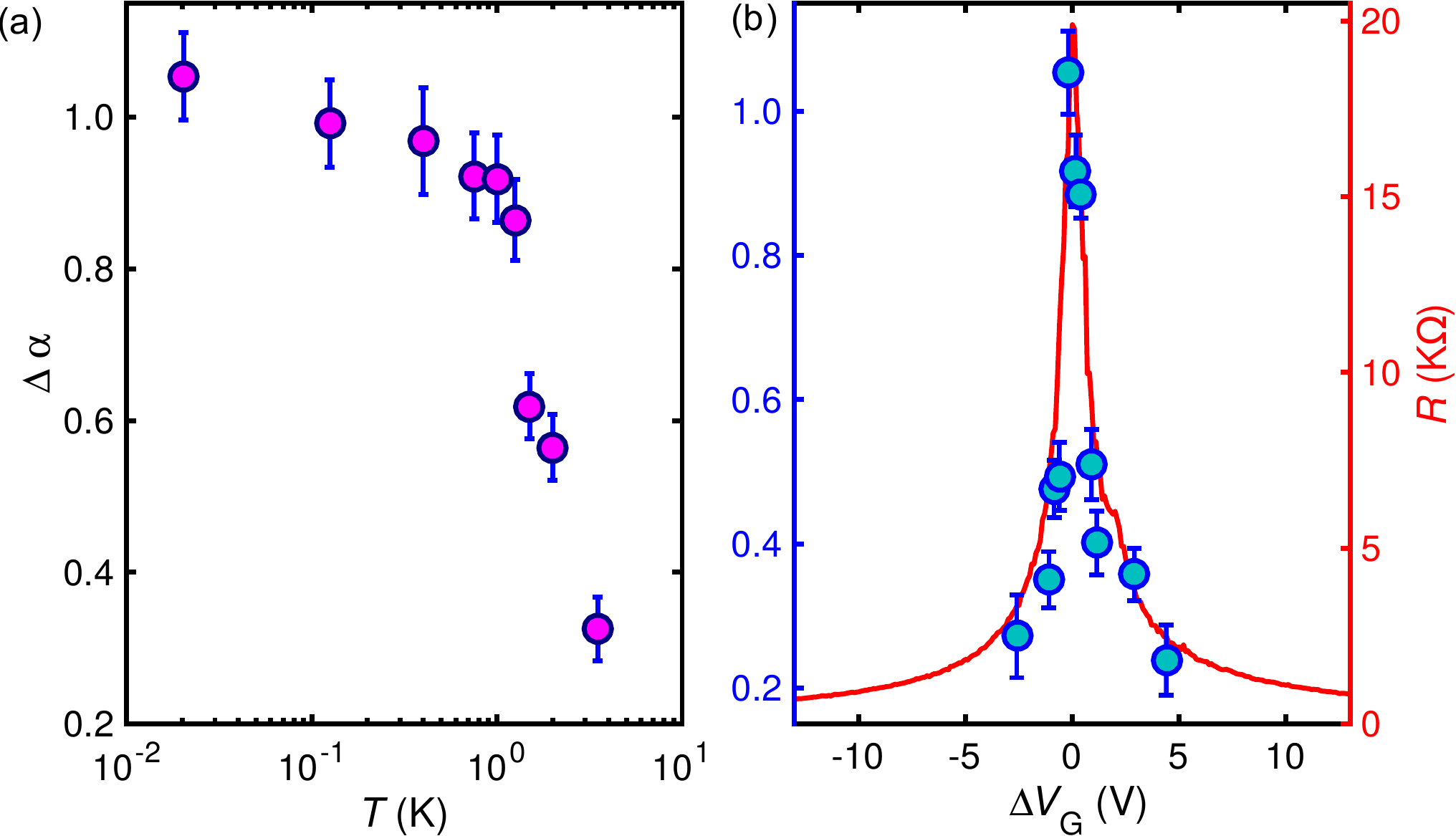}
    \small { \caption{  \textbf{Width of the multifractal spectrum.} (a) Blue filled circles (left axis)  show the  dependence of the width of the multifractal spectrum, $\Delta\alpha$ on $\Delta V_{\mathrm{G}}$; the data were obtained at  20~mK.  The red line shows the dependence of $R$ on $V_{\mathrm{G}}$, measured at 20~mK (right axis). It is clear from the plot that $\Delta\alpha$ is large only over a narrow range of gate voltage around $\Delta V_{\mathrm{G}}=0$~V, and sharply decreases as the system is driven away from the  Dirac point. 
        (b) Dependence of the width $\Delta\alpha$ on temperature, at $\Delta V_{\mathrm{G}}=0.2$~V; the data were obtained from the UCF plots shown in \ref*{fig:ucf_T}(a).
 The error-bars in (a) and (b) represent the standard deviation in $\Delta \alpha$.
        \label{fig:width} }}
  \end{center}
\end {figure*}

The most compelling explanation of the multifractal UCF we observe in our SLG samples is an incipient Anderson localization near the charge-neutrality point.  
Multifractality of the local amplitudes of critical eigenstates near Anderson localization has been studied, theoretically, in several quantum-condensed-matter systems~\cite{Mirlin2000259,PhysRevLett.111.066601, PhysRevB.87.125144,2053-1583-1-1-011009}. The multifractality of the eigenstates near the critical point directly affects the two-particle correlation function through the generalized diffusion coefficient~\cite{ANDP:ANDP2065080803,0295-5075-4-12-003}, which, in turn, affects the local current fluctuations in the system  via the Kubo formula. It is not obvious that this must be reflected in the (macroscopic) conductance, or its moments; however, it is plausible that near the critical point, the UCF may inherit multifractal behaviour from its counterpart in the eigenfunctions~\cite{PhysRevLett.87.014101}.  Indeed, there are several theoretical predictions of multifractality in transport coefficients including conductance jumps near the percolation threshold in random-resistor networks~\cite{PhysRevLett.54.1718, PhysRevB.43.3601}, conductance fluctuations in  quantum-Hall transitions~\cite{PhysRevLett.72.3582}, and the temperature dependence of the peak height of the conductance at the Anderson-localization transition~\cite{PhysRevLett.81.4696}.

Spectroscopic studies on single-layer on-substrate  graphene devices have revealed that the local potential fluctuations in this system are strongest when  $E_F$ is close to the Dirac point~\cite{PhysRevLett.116.126804, Jung2011}. This leads to electronic states that are   quasi-localized ~\cite{PhysRevLett.104.096804,  PhysRevLett.99.236801, RevModPhys.83.407}. 
Such  quasi-localized states have a high inverse participation ratio~\cite{PhysRevLett.96.036801} that can lead to the multifractality seen in our experiments. If this is  true, then the multifractality in the UCF should be largest near the Dirac point; and then fall off on either side of it. In \ref{fig:width} we show the dependence of $\Delta\alpha$ on $T$ and $\Delta V_{\mathrm{G}}$ for the device G28M6 [data  obtained for other devices are qualitatively similar]. We observe that   $\Delta\alpha$ is indeed largest near $\Delta V_{\mathrm{G}}=0$ and at low $T$, where the conductance of the device is of the order of $e^2/h$, and it sharply decreases  as  either $T$ or the magnitude of $\Delta V_{\mathrm{G}}$ increases. 
Similarly,  as $T$  is increased, thermal scattering increases quasiparticle dephasing, and eventually at high $T$, when $L_\upphi \sim L_{\mathrm{T}} \ll L$, quantum-interference effects are masked. From our observation that a large multifractality arises only when quantum-interference-induced charge-carrier localization is significant, we propose that an incipient Anderson localization near the Dirac point is the most plausible origin of multifractal UCF in SLG.

This interpretation of the multifractality of the UCF in single-layer graphene devices is based on previous theoretical predictions and analysis. We summarize our argument below:

Conductance fluctuations, as a function of the magnetic field,  have been shown to have a fractional fractal dimensions
 in some simple, one-dimensional, quantum systems, e.g., the kicked rotor~\cite{PhysRevLett.87.014101,PhysRevLett.84.63}. In these systems, such a fractional fractal dimension arises if one of the following conditions holds:
(1) The PDF $P(t)$, of the charge-carrier survival time $t$, has a power-law form $P(t) \propto t^{-\gamma}$ at large $t$ (as opposed to an exponential decay); (2) the energy correlation $C(\Delta E)$ of elements of the S-matrix exhibit power  laws (i.e., $C(\Delta E) \propto (\Delta E)^{-\gamma}$)~\cite{PhysRevB.54.10841,PhysRevLett.84.63, PhysRevLett.77.3885,PhysRevLett.68.3491}. Such survival probabilities are related to the conductance~\cite{PhysRevLett.84.63,PhysRevLett.87.014101}; and their multifractal behaviour has been explored~\cite{Facchini2007266}.

At the Anderson-localization transition, it is known that both the probability density function describing the diffusion of a wave-packet  and the two-particle correlation function decay  algebraically with a fractional power~\cite{ANDP:ANDP2065080803,0295-5075-4-12-003,nakayama2013fractal}. Hence, as in the case of the simple quantum systems mentioned above, we may expect multifractal fluctuations of the conductance at the Anderson-localization transition. However, to the best of our knowledge,   there are no exact, analytical results  that yield a one-to-one correspondence between the multifractality of a critical wavefunction and the multifractality of the magnetoconductance. Thus, we propose that the multifractality of the critical wavefunctions at the Anderson localization is the most plausible cause of the multifractality of the UCF we have observed.

\section{Conclusions}
In conclusion, we have uncovered and quantified the multifractal structure of mesoscopic conductance fluctuations in single-layer graphene devices. We speculate that our results are indicative of an incipient Anderson localization in this system. In particular, we quantify the multifractality of transport in a quantum condensed-matter system. There may well be multifractality in transport properties in systems other than graphene, and that multifractality is not unique to graphene. Our work provides a natural framework for studying the multifractality of such transport properties.

\section{Acknowledgements} 

We acknowledge discussions with A. D. Mirlin and S. Bhattacharjee. A.B. acknowledges financial support from Nanomission, DST, Govt. of India  project SR/NM/NS-35/2012; SERB, DST, Govt. of India  and Indo-French Centre for the Promotion of Advanced Recearch (CEFIPRA). K.R.A. thanks CSIR, MHRD, Govt. of India for financial support.  S.S.R. acknowledge financial support from the AIRBUS Group Corporate Foundation Chair in Mathematics of Complex Systems established in ICTS-TIFR, the support from  DST,  Govt. of India, project ECR/2015/000361,  and the Indo-French Center for Applied Mathematics (IFCAM). R.P. acknowledges DST, Govt. of India for support. N.P. thanks UGC, Govt. of India for financial support.

\section{Author contributions}

K.R.A and A.B. designed the experiment, fabricated the devices and carried out the measurements.  K.R.A., A.B., N.P. and S.S.R. carried out the data analysis. All authors discussed the results and wrote the paper.


\section*{References}


\begin{thebibliography}{70}%
\makeatletter
\providecommand \@ifxundefined [1]{%
 \@ifx{#1\undefined}
}%
\providecommand \@ifnum [1]{%
 \ifnum #1\expandafter \@firstoftwo
 \else \expandafter \@secondoftwo
 \fi
}%
\providecommand \@ifx [1]{%
 \ifx #1\expandafter \@firstoftwo
 \else \expandafter \@secondoftwo
 \fi
}%
\providecommand \natexlab [1]{#1}%
\providecommand \enquote  [1]{``#1''}%
\providecommand \bibnamefont  [1]{#1}%
\providecommand \bibfnamefont [1]{#1}%
\providecommand \citenamefont [1]{#1}%
\providecommand \href@noop [0]{\@secondoftwo}%
\providecommand \href [0]{\begingroup \@sanitize@url \@href}%
\providecommand \@href[1]{\@@startlink{#1}\@@href}%
\providecommand \@@href[1]{\endgroup#1\@@endlink}%
\providecommand \@sanitize@url [0]{\catcode `\\12\catcode `\$12\catcode
  `\&12\catcode `\#12\catcode `\^12\catcode `\_12\catcode `\%12\relax}%
\providecommand \@@startlink[1]{}%
\providecommand \@@endlink[0]{}%
\providecommand \url  [0]{\begingroup\@sanitize@url \@url }%
\providecommand \@url [1]{\endgroup\@href {#1}{\urlprefix }}%
\providecommand \urlprefix  [0]{URL }%
\providecommand \Eprint [0]{\href }%
\providecommand \doibase [0]{http://dx.doi.org/}%
\providecommand \selectlanguage [0]{\@gobble}%
\providecommand \bibinfo  [0]{\@secondoftwo}%
\providecommand \bibfield  [0]{\@secondoftwo}%
\providecommand \translation [1]{[#1]}%
\providecommand \BibitemOpen [0]{}%
\providecommand \bibitemStop [0]{}%
\providecommand \bibitemNoStop [0]{.\EOS\space}%
\providecommand \EOS [0]{\spacefactor3000\relax}%
\providecommand \BibitemShut  [1]{\csname bibitem#1\endcsname}%
\let\auto@bib@innerbib\@empty
\bibitem [{\citenamefont {Evers}\ and\ \citenamefont
  {Mirlin}(2008)}]{RevModPhys.80.1355}%
  \BibitemOpen
  \bibfield  {author} {\bibinfo {author} {\bibfnamefont {F.}~\bibnamefont
  {Evers}}\ and\ \bibinfo {author} {\bibfnamefont {A.~D.}\ \bibnamefont
  {Mirlin}},\ }\href {\doibase 10.1103/RevModPhys.80.1355} {\bibfield
  {journal} {\bibinfo  {journal} {Rev. Mod. Phys.}\ }\textbf {\bibinfo {volume}
  {80}},\ \bibinfo {pages} {1355} (\bibinfo {year} {2008})}\BibitemShut
  {NoStop}%
\bibitem [{\citenamefont {Ludwig}\ \emph {et~al.}(1994)\citenamefont {Ludwig},
  \citenamefont {Fisher}, \citenamefont {Shankar},\ and\ \citenamefont
  {Grinstein}}]{PhysRevB.50.7526}%
  \BibitemOpen
  \bibfield  {author} {\bibinfo {author} {\bibfnamefont {A.~W.~W.}\
  \bibnamefont {Ludwig}}, \bibinfo {author} {\bibfnamefont {M.~P.~A.}\
  \bibnamefont {Fisher}}, \bibinfo {author} {\bibfnamefont {R.}~\bibnamefont
  {Shankar}}, \ and\ \bibinfo {author} {\bibfnamefont {G.}~\bibnamefont
  {Grinstein}},\ }\href {\doibase 10.1103/PhysRevB.50.7526} {\bibfield
  {journal} {\bibinfo  {journal} {Phys. Rev. B}\ }\textbf {\bibinfo {volume}
  {50}},\ \bibinfo {pages} {7526} (\bibinfo {year} {1994})}\BibitemShut
  {NoStop}%
\bibitem [{\citenamefont {Cheianov}\ and\ \citenamefont
  {Falko}(2006)}]{PhysRevB.74.041403}%
  \BibitemOpen
  \bibfield  {author} {\bibinfo {author} {\bibfnamefont {V.~V.}\ \bibnamefont
  {Cheianov}}\ and\ \bibinfo {author} {\bibfnamefont {V.~I.}\ \bibnamefont
  {Falko}},\ }\href {\doibase 10.1103/PhysRevB.74.041403} {\bibfield  {journal}
  {\bibinfo  {journal} {Phys. Rev. B}\ }\textbf {\bibinfo {volume} {74}},\
  \bibinfo {pages} {041403} (\bibinfo {year} {2006})}\BibitemShut {NoStop}%
\bibitem [{\citenamefont {Aleiner}\ and\ \citenamefont
  {Efetov}(2006)}]{PhysRevLett.97.236801}%
  \BibitemOpen
  \bibfield  {author} {\bibinfo {author} {\bibfnamefont {I.~L.}\ \bibnamefont
  {Aleiner}}\ and\ \bibinfo {author} {\bibfnamefont {K.~B.}\ \bibnamefont
  {Efetov}},\ }\href {\doibase 10.1103/PhysRevLett.97.236801} {\bibfield
  {journal} {\bibinfo  {journal} {Phys. Rev. Lett.}\ }\textbf {\bibinfo
  {volume} {97}},\ \bibinfo {pages} {236801} (\bibinfo {year}
  {2006})}\BibitemShut {NoStop}%
\bibitem [{\citenamefont {Katsnelson}\ \emph {et~al.}(2006)\citenamefont
  {Katsnelson}, \citenamefont {Novoselov},\ and\ \citenamefont
  {Geim}}]{Katsnelson2006}%
  \BibitemOpen
  \bibfield  {author} {\bibinfo {author} {\bibfnamefont {M.~I.}\ \bibnamefont
  {Katsnelson}}, \bibinfo {author} {\bibfnamefont {K.~S.}\ \bibnamefont
  {Novoselov}}, \ and\ \bibinfo {author} {\bibfnamefont {A.~K.}\ \bibnamefont
  {Geim}},\ }\href {\doibase 10.1038/nphys384} {\bibfield  {journal} {\bibinfo
  {journal} {Nat. Phys.}\ }\textbf {\bibinfo {volume} {2}},\ \bibinfo {pages}
  {620} (\bibinfo {year} {2006})}\BibitemShut {NoStop}%
\bibitem [{\citenamefont {Pereira}\ \emph {et~al.}(2006)\citenamefont
  {Pereira}, \citenamefont {Guinea}, \citenamefont {Lopes~dos Santos},
  \citenamefont {Peres},\ and\ \citenamefont
  {Castro~Neto}}]{PhysRevLett.96.036801}%
  \BibitemOpen
  \bibfield  {author} {\bibinfo {author} {\bibfnamefont {V.~M.}\ \bibnamefont
  {Pereira}}, \bibinfo {author} {\bibfnamefont {F.}~\bibnamefont {Guinea}},
  \bibinfo {author} {\bibfnamefont {J.~M.~B.}\ \bibnamefont {Lopes~dos
  Santos}}, \bibinfo {author} {\bibfnamefont {N.~M.~R.}\ \bibnamefont {Peres}},
  \ and\ \bibinfo {author} {\bibfnamefont {A.~H.}\ \bibnamefont
  {Castro~Neto}},\ }\href {\doibase 10.1103/PhysRevLett.96.036801} {\bibfield
  {journal} {\bibinfo  {journal} {Phys. Rev. Lett.}\ }\textbf {\bibinfo
  {volume} {96}},\ \bibinfo {pages} {036801} (\bibinfo {year}
  {2006})}\BibitemShut {NoStop}%
\bibitem [{\citenamefont {Samaddar}\ \emph {et~al.}(2016)\citenamefont
  {Samaddar}, \citenamefont {Yudhistira}, \citenamefont {Adam}, \citenamefont
  {Courtois},\ and\ \citenamefont {Winkelmann}}]{PhysRevLett.116.126804}%
  \BibitemOpen
  \bibfield  {author} {\bibinfo {author} {\bibfnamefont {S.}~\bibnamefont
  {Samaddar}}, \bibinfo {author} {\bibfnamefont {I.}~\bibnamefont
  {Yudhistira}}, \bibinfo {author} {\bibfnamefont {S.}~\bibnamefont {Adam}},
  \bibinfo {author} {\bibfnamefont {H.}~\bibnamefont {Courtois}}, \ and\
  \bibinfo {author} {\bibfnamefont {C.~B.}\ \bibnamefont {Winkelmann}},\ }\href
  {\doibase 10.1103/PhysRevLett.116.126804} {\bibfield  {journal} {\bibinfo
  {journal} {Phys. Rev. Lett.}\ }\textbf {\bibinfo {volume} {116}},\ \bibinfo
  {pages} {126804} (\bibinfo {year} {2016})}\BibitemShut {NoStop}%
\bibitem [{\citenamefont {Bostwick}\ \emph {et~al.}(2009)\citenamefont
  {Bostwick}, \citenamefont {McChesney}, \citenamefont {Emtsev}, \citenamefont
  {Seyller}, \citenamefont {Horn}, \citenamefont {Kevan},\ and\ \citenamefont
  {Rotenberg}}]{PhysRevLett.103.056404}%
  \BibitemOpen
  \bibfield  {author} {\bibinfo {author} {\bibfnamefont {A.}~\bibnamefont
  {Bostwick}}, \bibinfo {author} {\bibfnamefont {J.~L.}\ \bibnamefont
  {McChesney}}, \bibinfo {author} {\bibfnamefont {K.~V.}\ \bibnamefont
  {Emtsev}}, \bibinfo {author} {\bibfnamefont {T.}~\bibnamefont {Seyller}},
  \bibinfo {author} {\bibfnamefont {K.}~\bibnamefont {Horn}}, \bibinfo {author}
  {\bibfnamefont {S.~D.}\ \bibnamefont {Kevan}}, \ and\ \bibinfo {author}
  {\bibfnamefont {E.}~\bibnamefont {Rotenberg}},\ }\href {\doibase
  10.1103/PhysRevLett.103.056404} {\bibfield  {journal} {\bibinfo  {journal}
  {Phys. Rev. Lett.}\ }\textbf {\bibinfo {volume} {103}},\ \bibinfo {pages}
  {056404} (\bibinfo {year} {2009})}\BibitemShut {NoStop}%
\bibitem [{\citenamefont {Jung}\ \emph {et~al.}(2011)\citenamefont {Jung},
  \citenamefont {Rutter}, \citenamefont {Klimov}, \citenamefont {Newell},
  \citenamefont {Calizo}, \citenamefont {Hight-Walker}, \citenamefont
  {Zhitenev},\ and\ \citenamefont {Stroscio}}]{Jung2011}%
  \BibitemOpen
  \bibfield  {author} {\bibinfo {author} {\bibfnamefont {S.}~\bibnamefont
  {Jung}}, \bibinfo {author} {\bibfnamefont {G.~M.}\ \bibnamefont {Rutter}},
  \bibinfo {author} {\bibfnamefont {N.~N.}\ \bibnamefont {Klimov}}, \bibinfo
  {author} {\bibfnamefont {D.~B.}\ \bibnamefont {Newell}}, \bibinfo {author}
  {\bibfnamefont {I.}~\bibnamefont {Calizo}}, \bibinfo {author} {\bibfnamefont
  {A.~R.}\ \bibnamefont {Hight-Walker}}, \bibinfo {author} {\bibfnamefont
  {N.~B.}\ \bibnamefont {Zhitenev}}, \ and\ \bibinfo {author} {\bibfnamefont
  {J.~A.}\ \bibnamefont {Stroscio}},\ }\href {\doibase 10.1038/nphys1866}
  {\bibfield  {journal} {\bibinfo  {journal} {Nat. Phys.}\ }\textbf {\bibinfo
  {volume} {7}},\ \bibinfo {pages} {245} (\bibinfo {year} {2011})}\BibitemShut
  {NoStop}%
\bibitem [{\citenamefont {Ponomarenko}\ \emph {et~al.}(2011)\citenamefont
  {Ponomarenko}, \citenamefont {Geim}, \citenamefont {Zhukov}, \citenamefont
  {Jalil}, \citenamefont {Morozov}, \citenamefont {Novoselov}, \citenamefont
  {Grigorieva}, \citenamefont {Hill}, \citenamefont {Cheianov}, \citenamefont
  {Fal/'ko}, \citenamefont {Watanabe}, \citenamefont {Taniguchi},\ and\
  \citenamefont {Gorbachev}}]{Ponomarenko2011}%
  \BibitemOpen
  \bibfield  {author} {\bibinfo {author} {\bibfnamefont {L.~A.}\ \bibnamefont
  {Ponomarenko}}, \bibinfo {author} {\bibfnamefont {A.~K.}\ \bibnamefont
  {Geim}}, \bibinfo {author} {\bibfnamefont {A.~A.}\ \bibnamefont {Zhukov}},
  \bibinfo {author} {\bibfnamefont {R.}~\bibnamefont {Jalil}}, \bibinfo
  {author} {\bibfnamefont {S.~V.}\ \bibnamefont {Morozov}}, \bibinfo {author}
  {\bibfnamefont {K.~S.}\ \bibnamefont {Novoselov}}, \bibinfo {author}
  {\bibfnamefont {I.~V.}\ \bibnamefont {Grigorieva}}, \bibinfo {author}
  {\bibfnamefont {E.~H.}\ \bibnamefont {Hill}}, \bibinfo {author}
  {\bibfnamefont {V.~V.}\ \bibnamefont {Cheianov}}, \bibinfo {author}
  {\bibfnamefont {V.~I.}\ \bibnamefont {Fal/'ko}}, \bibinfo {author}
  {\bibfnamefont {K.}~\bibnamefont {Watanabe}}, \bibinfo {author}
  {\bibfnamefont {T.}~\bibnamefont {Taniguchi}}, \ and\ \bibinfo {author}
  {\bibfnamefont {R.~V.}\ \bibnamefont {Gorbachev}},\ }\href {\doibase
  10.1038/nphys2114} {\bibfield  {journal} {\bibinfo  {journal} {Nat. Phys.}\
  }\textbf {\bibinfo {volume} {7}},\ \bibinfo {pages} {958} (\bibinfo {year}
  {2011})}\BibitemShut {NoStop}%
\bibitem [{\citenamefont {Mandelbrot}(1982)}]{mandelbrot19827he}%
  \BibitemOpen
  \bibfield  {author} {\bibinfo {author} {\bibfnamefont {B.~B.}\ \bibnamefont
  {Mandelbrot}},\ }\href@noop {} {\emph {\bibinfo {title} {The Fractal Geometry
  of Nature}}}\ (\bibinfo  {publisher} {Freeman: San Francisco},\ \bibinfo
  {year} {1982})\BibitemShut {NoStop}%
\bibitem [{\citenamefont {Ivanov}\ \emph {et~al.}(1999)\citenamefont {Ivanov},
  \citenamefont {Amaral}, \citenamefont {Goldberger}, \citenamefont {Havlin},
  \citenamefont {Rosenblum}, \citenamefont {Struzik},\ and\ \citenamefont
  {Stanley}}]{Ivanov1999}%
  \BibitemOpen
  \bibfield  {author} {\bibinfo {author} {\bibfnamefont {P.~C.}\ \bibnamefont
  {Ivanov}}, \bibinfo {author} {\bibfnamefont {L.~A.~N.}\ \bibnamefont
  {Amaral}}, \bibinfo {author} {\bibfnamefont {A.~L.}\ \bibnamefont
  {Goldberger}}, \bibinfo {author} {\bibfnamefont {S.}~\bibnamefont {Havlin}},
  \bibinfo {author} {\bibfnamefont {M.~G.}\ \bibnamefont {Rosenblum}}, \bibinfo
  {author} {\bibfnamefont {Z.~R.}\ \bibnamefont {Struzik}}, \ and\ \bibinfo
  {author} {\bibfnamefont {H.~E.}\ \bibnamefont {Stanley}},\ }\href
  {http://dx.doi.org/10.1038/20924} {\bibfield  {journal} {\bibinfo  {journal}
  {Nature}\ }\textbf {\bibinfo {volume} {399}},\ \bibinfo {pages} {461}
  (\bibinfo {year} {1999})}\BibitemShut {NoStop}%
\bibitem [{\citenamefont {Lawrence}\ \emph {et~al.}(1993)\citenamefont
  {Lawrence}, \citenamefont {Ruzmaikin},\ and\ \citenamefont
  {Cadavid}}]{lawrence1993multifractal}%
  \BibitemOpen
  \bibfield  {author} {\bibinfo {author} {\bibfnamefont {J.}~\bibnamefont
  {Lawrence}}, \bibinfo {author} {\bibfnamefont {A.}~\bibnamefont {Ruzmaikin}},
  \ and\ \bibinfo {author} {\bibfnamefont {A.}~\bibnamefont {Cadavid}},\
  }\href@noop {} {\bibfield  {journal} {\bibinfo  {journal} {The Astrophysical
  Journal}\ }\textbf {\bibinfo {volume} {417}},\ \bibinfo {pages} {805}
  (\bibinfo {year} {1993})}\BibitemShut {NoStop}%
\bibitem [{\citenamefont {Oudjemia}\ \emph {et~al.}(2013)\citenamefont
  {Oudjemia}, \citenamefont {Girault}, \citenamefont {e.~Derguini},\ and\
  \citenamefont {Haddab}}]{6602370}%
  \BibitemOpen
  \bibfield  {author} {\bibinfo {author} {\bibfnamefont {S.}~\bibnamefont
  {Oudjemia}}, \bibinfo {author} {\bibfnamefont {J.~M.}\ \bibnamefont
  {Girault}}, \bibinfo {author} {\bibfnamefont {N.}~\bibnamefont
  {e.~Derguini}}, \ and\ \bibinfo {author} {\bibfnamefont {S.}~\bibnamefont
  {Haddab}},\ }in\ \href {\doibase 10.1109/WoSSPA.2013.6602370} {\emph
  {\bibinfo {booktitle} {8th International Workshop on Systems, Signal
  Processing and their  Applications }}}\ (\bibinfo {year} {2013})\ pp.\
  \bibinfo {pages} {244--249}\BibitemShut {NoStop}%
\bibitem [{\citenamefont {Frisch}(1995)}]{frisch1995turbulence}%
  \BibitemOpen
  \bibfield  {author} {\bibinfo {author} {\bibfnamefont {U.}~\bibnamefont
  {Frisch}},\ }\href@noop {} {\emph {\bibinfo {title} {Turbulence: the legacy
  of AN Kolmogorov}}}\ (\bibinfo  {publisher} {Cambridge university press},\
  \bibinfo {year} {1995})\BibitemShut {NoStop}%
\bibitem [{\citenamefont {Benzi}\ \emph {et~al.}(1984)\citenamefont {Benzi},
  \citenamefont {Paladin}, \citenamefont {Parisi},\ and\ \citenamefont
  {Vulpiani}}]{0305-4470-17-18-021}%
  \BibitemOpen
  \bibfield  {author} {\bibinfo {author} {\bibfnamefont {R.}~\bibnamefont
  {Benzi}}, \bibinfo {author} {\bibfnamefont {G.}~\bibnamefont {Paladin}},
  \bibinfo {author} {\bibfnamefont {G.}~\bibnamefont {Parisi}}, \ and\ \bibinfo
  {author} {\bibfnamefont {A.}~\bibnamefont {Vulpiani}},\ }\href
  {http://stacks.iop.org/0305-4470/17/i=18/a=021} {\bibfield  {journal}
  {\bibinfo  {journal} {J. Phys. A: Math. Gen.}\ }\textbf {\bibinfo {volume}
  {17}},\ \bibinfo {pages} {3521} (\bibinfo {year} {1984})}\BibitemShut
  {NoStop}%
\bibitem [{\citenamefont {Castellani}\ and\ \citenamefont
  {Peliti}(1986)}]{0305-4470-19-8-004}%
  \BibitemOpen
  \bibfield  {author} {\bibinfo {author} {\bibfnamefont {C.}~\bibnamefont
  {Castellani}}\ and\ \bibinfo {author} {\bibfnamefont {L.}~\bibnamefont
  {Peliti}},\ }\href {http://stacks.iop.org/0305-4470/19/i=8/a=004} {\bibfield
  {journal} {\bibinfo  {journal} {J. Phys. A: Math. Gen.}\ }\textbf {\bibinfo
  {volume} {19}},\ \bibinfo {pages} {L429} (\bibinfo {year}
  {1986})}\BibitemShut {NoStop}%
\bibitem [{\citenamefont {Janssen}(1998)}]{JANSSEN19981}%
  \BibitemOpen
  \bibfield  {author} {\bibinfo {author} {\bibfnamefont {M.}~\bibnamefont
  {Janssen}},\ }\href {\doibase
  http://dx.doi.org/10.1016/S0370-1573(97)00050-1} {\bibfield  {journal}
  {\bibinfo  {journal} {Phys. Rep.}\ }\textbf {\bibinfo {volume} {295}},\
  \bibinfo {pages} {1 } (\bibinfo {year} {1998})}\BibitemShut {NoStop}%
\bibitem [{\citenamefont {Mirlin}(2000)}]{Mirlin2000259}%
  \BibitemOpen
  \bibfield  {author} {\bibinfo {author} {\bibfnamefont {A.~D.}\ \bibnamefont
  {Mirlin}},\ }\href {\doibase http://dx.doi.org/10.1016/S0370-1573(99)00091-5}
  {\bibfield  {journal} {\bibinfo  {journal} {Phys. Rep.}\ }\textbf {\bibinfo
  {volume} {326}},\ \bibinfo {pages} {259 } (\bibinfo {year}
  {2000})}\BibitemShut {NoStop}%
\bibitem [{\citenamefont {Grussbach}\ and\ \citenamefont
  {Schreiber}(1995)}]{PhysRevB.51.663}%
  \BibitemOpen
  \bibfield  {author} {\bibinfo {author} {\bibfnamefont {H.}~\bibnamefont
  {Grussbach}}\ and\ \bibinfo {author} {\bibfnamefont {M.}~\bibnamefont
  {Schreiber}},\ }\href {\doibase 10.1103/PhysRevB.51.663} {\bibfield
  {journal} {\bibinfo  {journal} {Phys. Rev. B}\ }\textbf {\bibinfo {volume}
  {51}},\ \bibinfo {pages} {663} (\bibinfo {year} {1995})}\BibitemShut
  {NoStop}%
\bibitem [{\citenamefont {Burmistrov}\ \emph {et~al.}(2015)\citenamefont
  {Burmistrov}, \citenamefont {Gornyi},\ and\ \citenamefont
  {Mirlin}}]{PhysRevB.91.085427}%
  \BibitemOpen
  \bibfield  {author} {\bibinfo {author} {\bibfnamefont {I.~S.}\ \bibnamefont
  {Burmistrov}}, \bibinfo {author} {\bibfnamefont {I.~V.}\ \bibnamefont
  {Gornyi}}, \ and\ \bibinfo {author} {\bibfnamefont {A.~D.}\ \bibnamefont
  {Mirlin}},\ }\href {\doibase 10.1103/PhysRevB.91.085427} {\bibfield
  {journal} {\bibinfo  {journal} {Phys. Rev. B}\ }\textbf {\bibinfo {volume}
  {91}},\ \bibinfo {pages} {085427} (\bibinfo {year} {2015})}\BibitemShut
  {NoStop}%
\bibitem [{\citenamefont {Burmistrov}\ \emph {et~al.}(2013)\citenamefont
  {Burmistrov}, \citenamefont {Gornyi},\ and\ \citenamefont
  {Mirlin}}]{PhysRevLett.111.066601}%
  \BibitemOpen
  \bibfield  {author} {\bibinfo {author} {\bibfnamefont {I.~S.}\ \bibnamefont
  {Burmistrov}}, \bibinfo {author} {\bibfnamefont {I.~V.}\ \bibnamefont
  {Gornyi}}, \ and\ \bibinfo {author} {\bibfnamefont {A.~D.}\ \bibnamefont
  {Mirlin}},\ }\href {\doibase 10.1103/PhysRevLett.111.066601} {\bibfield
  {journal} {\bibinfo  {journal} {Phys. Rev. Lett.}\ }\textbf {\bibinfo
  {volume} {111}},\ \bibinfo {pages} {066601} (\bibinfo {year}
  {2013})}\BibitemShut {NoStop}%
\bibitem [{\citenamefont {Gruzberg}\ \emph {et~al.}(2013)\citenamefont
  {Gruzberg}, \citenamefont {Mirlin},\ and\ \citenamefont
  {Zirnbauer}}]{PhysRevB.87.125144}%
  \BibitemOpen
  \bibfield  {author} {\bibinfo {author} {\bibfnamefont {I.~A.}\ \bibnamefont
  {Gruzberg}}, \bibinfo {author} {\bibfnamefont {A.~D.}\ \bibnamefont
  {Mirlin}}, \ and\ \bibinfo {author} {\bibfnamefont {M.~R.}\ \bibnamefont
  {Zirnbauer}},\ }\href {\doibase 10.1103/PhysRevB.87.125144} {\bibfield
  {journal} {\bibinfo  {journal} {Phys. Rev. B}\ }\textbf {\bibinfo {volume}
  {87}},\ \bibinfo {pages} {125144} (\bibinfo {year} {2013})}\BibitemShut
  {NoStop}%
\bibitem [{\citenamefont {Barrios-Vargas}\ and\ \citenamefont
  {Naumis}(2014)}]{2053-1583-1-1-011009}%
  \BibitemOpen
  \bibfield  {author} {\bibinfo {author} {\bibfnamefont {J.~E.}\ \bibnamefont
  {Barrios-Vargas}}\ and\ \bibinfo {author} {\bibfnamefont {G.~G.}\
  \bibnamefont {Naumis}},\ }\href
  {http://stacks.iop.org/2053-1583/1/i=1/a=011009} {\bibfield  {journal}
  {\bibinfo  {journal} {2D Mater.}\ }\textbf {\bibinfo {volume} {1}},\ \bibinfo
  {pages} {011009} (\bibinfo {year} {2014})}\BibitemShut {NoStop}%
\bibitem [{\citenamefont {Facchini}\ \emph {et~al.}(2007)\citenamefont
  {Facchini}, \citenamefont {Wimberger},\ and\ \citenamefont
  {Tomadin}}]{Facchini2007266}%
  \BibitemOpen
  \bibfield  {author} {\bibinfo {author} {\bibfnamefont {A.}~\bibnamefont
  {Facchini}}, \bibinfo {author} {\bibfnamefont {S.}~\bibnamefont {Wimberger}},
  \ and\ \bibinfo {author} {\bibfnamefont {A.}~\bibnamefont {Tomadin}},\ }\href
  {\doibase http://dx.doi.org/10.1016/j.physa.2006.10.012} {\bibfield
  {journal} {\bibinfo  {journal} {Physica A}\ }\textbf {\bibinfo {volume}
  {376}},\ \bibinfo {pages} {266 } (\bibinfo {year} {2007})}\BibitemShut
  {NoStop}%
\bibitem [{\citenamefont {Monthus}\ and\ \citenamefont
  {Garel}(2009)}]{PhysRevB.79.205120}%
  \BibitemOpen
  \bibfield  {author} {\bibinfo {author} {\bibfnamefont {C.}~\bibnamefont
  {Monthus}}\ and\ \bibinfo {author} {\bibfnamefont {T.}~\bibnamefont
  {Garel}},\ }\href {\doibase 10.1103/PhysRevB.79.205120} {\bibfield  {journal}
  {\bibinfo  {journal} {Phys. Rev. B}\ }\textbf {\bibinfo {volume} {79}},\
  \bibinfo {pages} {205120} (\bibinfo {year} {2009})}\BibitemShut {NoStop}%
\bibitem [{\citenamefont {Rammal}\ \emph {et~al.}(1985)\citenamefont {Rammal},
  \citenamefont {Tannous}, \citenamefont {Breton},\ and\ \citenamefont
  {Tremblay}}]{PhysRevLett.54.1718}%
  \BibitemOpen
  \bibfield  {author} {\bibinfo {author} {\bibfnamefont {R.}~\bibnamefont
  {Rammal}}, \bibinfo {author} {\bibfnamefont {C.}~\bibnamefont {Tannous}},
  \bibinfo {author} {\bibfnamefont {P.}~\bibnamefont {Breton}}, \ and\ \bibinfo
  {author} {\bibfnamefont {A.~M.~S.}\ \bibnamefont {Tremblay}},\ }\href
  {\doibase 10.1103/PhysRevLett.54.1718} {\bibfield  {journal} {\bibinfo
  {journal} {Phys. Rev. Lett.}\ }\textbf {\bibinfo {volume} {54}},\ \bibinfo
  {pages} {1718} (\bibinfo {year} {1985})}\BibitemShut {NoStop}%
\bibitem [{\citenamefont {Roux}\ \emph {et~al.}(1991)\citenamefont {Roux},
  \citenamefont {Hansen},\ and\ \citenamefont {Hinrichsen}}]{PhysRevB.43.3601}%
  \BibitemOpen
  \bibfield  {author} {\bibinfo {author} {\bibfnamefont {S.}~\bibnamefont
  {Roux}}, \bibinfo {author} {\bibfnamefont {A.}~\bibnamefont {Hansen}}, \ and\
  \bibinfo {author} {\bibfnamefont {E.~L.}\ \bibnamefont {Hinrichsen}},\ }\href
  {\doibase 10.1103/PhysRevB.43.3601} {\bibfield  {journal} {\bibinfo
  {journal} {Phys. Rev. B}\ }\textbf {\bibinfo {volume} {43}},\ \bibinfo
  {pages} {3601} (\bibinfo {year} {1991})}\BibitemShut {NoStop}%
\bibitem [{\citenamefont {Brandes}\ \emph {et~al.}(1994)\citenamefont
  {Brandes}, \citenamefont {Schweitzer},\ and\ \citenamefont
  {Kramer}}]{PhysRevLett.72.3582}%
  \BibitemOpen
  \bibfield  {author} {\bibinfo {author} {\bibfnamefont {T.}~\bibnamefont
  {Brandes}}, \bibinfo {author} {\bibfnamefont {L.}~\bibnamefont {Schweitzer}},
  \ and\ \bibinfo {author} {\bibfnamefont {B.}~\bibnamefont {Kramer}},\ }\href
  {\doibase 10.1103/PhysRevLett.72.3582} {\bibfield  {journal} {\bibinfo
  {journal} {Phys. Rev. Lett.}\ }\textbf {\bibinfo {volume} {72}},\ \bibinfo
  {pages} {3582} (\bibinfo {year} {1994})}\BibitemShut {NoStop}%
\bibitem [{\citenamefont {Polyakov}(1998)}]{PhysRevLett.81.4696}%
  \BibitemOpen
  \bibfield  {author} {\bibinfo {author} {\bibfnamefont {D.~G.}\ \bibnamefont
  {Polyakov}},\ }\href {\doibase 10.1103/PhysRevLett.81.4696} {\bibfield
  {journal} {\bibinfo  {journal} {Phys. Rev. Lett.}\ }\textbf {\bibinfo
  {volume} {81}},\ \bibinfo {pages} {4696} (\bibinfo {year}
  {1998})}\BibitemShut {NoStop}%
\bibitem [{\citenamefont {Hegger}\ \emph {et~al.}(1996)\citenamefont {Hegger},
  \citenamefont {Huckestein}, \citenamefont {Hecker}, \citenamefont {Janssen},
  \citenamefont {Freimuth}, \citenamefont {Reckziegel},\ and\ \citenamefont
  {Tuzinski}}]{PhysRevLett.77.3885}%
  \BibitemOpen
  \bibfield  {author} {\bibinfo {author} {\bibfnamefont {H.}~\bibnamefont
  {Hegger}}, \bibinfo {author} {\bibfnamefont {B.}~\bibnamefont {Huckestein}},
  \bibinfo {author} {\bibfnamefont {K.}~\bibnamefont {Hecker}}, \bibinfo
  {author} {\bibfnamefont {M.}~\bibnamefont {Janssen}}, \bibinfo {author}
  {\bibfnamefont {A.}~\bibnamefont {Freimuth}}, \bibinfo {author}
  {\bibfnamefont {G.}~\bibnamefont {Reckziegel}}, \ and\ \bibinfo {author}
  {\bibfnamefont {R.}~\bibnamefont {Tuzinski}},\ }\href {\doibase
  10.1103/PhysRevLett.77.3885} {\bibfield  {journal} {\bibinfo  {journal}
  {Phys. Rev. Lett.}\ }\textbf {\bibinfo {volume} {77}},\ \bibinfo {pages}
  {3885} (\bibinfo {year} {1996})}\BibitemShut {NoStop}%
\bibitem [{\citenamefont {Micolich}\ \emph {et~al.}(2001)\citenamefont
  {Micolich}, \citenamefont {Taylor}, \citenamefont {Davies}, \citenamefont
  {Bird}, \citenamefont {Newbury}, \citenamefont {Fromhold}, \citenamefont
  {Ehlert}, \citenamefont {Linke}, \citenamefont {Macks}, \citenamefont
  {Tribe}, \citenamefont {Linfield}, \citenamefont {Ritchie}, \citenamefont
  {Cooper}, \citenamefont {Aoyagi},\ and\ \citenamefont
  {Wilkinson}}]{PhysRevLett.87.036802}%
  \BibitemOpen
  \bibfield  {author} {\bibinfo {author} {\bibfnamefont {A.~P.}\ \bibnamefont
  {Micolich}}, \bibinfo {author} {\bibfnamefont {R.~P.}\ \bibnamefont
  {Taylor}}, \bibinfo {author} {\bibfnamefont {A.~G.}\ \bibnamefont {Davies}},
  \bibinfo {author} {\bibfnamefont {J.~P.}\ \bibnamefont {Bird}}, \bibinfo
  {author} {\bibfnamefont {R.}~\bibnamefont {Newbury}}, \bibinfo {author}
  {\bibfnamefont {T.~M.}\ \bibnamefont {Fromhold}}, \bibinfo {author}
  {\bibfnamefont {A.}~\bibnamefont {Ehlert}}, \bibinfo {author} {\bibfnamefont
  {H.}~\bibnamefont {Linke}}, \bibinfo {author} {\bibfnamefont {L.~D.}\
  \bibnamefont {Macks}}, \bibinfo {author} {\bibfnamefont {W.~R.}\ \bibnamefont
  {Tribe}}, \bibinfo {author} {\bibfnamefont {E.~H.}\ \bibnamefont {Linfield}},
  \bibinfo {author} {\bibfnamefont {D.~A.}\ \bibnamefont {Ritchie}}, \bibinfo
  {author} {\bibfnamefont {J.}~\bibnamefont {Cooper}}, \bibinfo {author}
  {\bibfnamefont {Y.}~\bibnamefont {Aoyagi}}, \ and\ \bibinfo {author}
  {\bibfnamefont {P.~B.}\ \bibnamefont {Wilkinson}},\ }\href {\doibase
  10.1103/PhysRevLett.87.036802} {\bibfield  {journal} {\bibinfo  {journal}
  {Phys. Rev. Lett.}\ }\textbf {\bibinfo {volume} {87}},\ \bibinfo {pages}
  {036802} (\bibinfo {year} {2001})}\BibitemShut {NoStop}%
\bibitem [{\citenamefont {Sachrajda}\ \emph {et~al.}(1998)\citenamefont
  {Sachrajda}, \citenamefont {Ketzmerick}, \citenamefont {Gould}, \citenamefont
  {Feng}, \citenamefont {Kelly}, \citenamefont {Delage},\ and\ \citenamefont
  {Wasilewski}}]{PhysRevLett.80.1948}%
  \BibitemOpen
  \bibfield  {author} {\bibinfo {author} {\bibfnamefont {A.~S.}\ \bibnamefont
  {Sachrajda}}, \bibinfo {author} {\bibfnamefont {R.}~\bibnamefont
  {Ketzmerick}}, \bibinfo {author} {\bibfnamefont {C.}~\bibnamefont {Gould}},
  \bibinfo {author} {\bibfnamefont {Y.}~\bibnamefont {Feng}}, \bibinfo {author}
  {\bibfnamefont {P.~J.}\ \bibnamefont {Kelly}}, \bibinfo {author}
  {\bibfnamefont {A.}~\bibnamefont {Delage}}, \ and\ \bibinfo {author}
  {\bibfnamefont {Z.}~\bibnamefont {Wasilewski}},\ }\href {\doibase
  10.1103/PhysRevLett.80.1948} {\bibfield  {journal} {\bibinfo  {journal}
  {Phys. Rev. Lett.}\ }\textbf {\bibinfo {volume} {80}},\ \bibinfo {pages}
  {1948} (\bibinfo {year} {1998})}\BibitemShut {NoStop}%
\bibitem [{\citenamefont {Ketzmerick}(1996)}]{PhysRevB.54.10841}%
  \BibitemOpen
  \bibfield  {author} {\bibinfo {author} {\bibfnamefont {R.}~\bibnamefont
  {Ketzmerick}},\ }\href {\doibase 10.1103/PhysRevB.54.10841} {\bibfield
  {journal} {\bibinfo  {journal} {Phys. Rev. B}\ }\textbf {\bibinfo {volume}
  {54}},\ \bibinfo {pages} {10841} (\bibinfo {year} {1996})}\BibitemShut
  {NoStop}%
\bibitem [{\citenamefont {Ujiie}\ \emph {et~al.}(2008)\citenamefont {Ujiie},
  \citenamefont {Yumoto}, \citenamefont {Morimoto}, \citenamefont {Aoki},\ and\
  \citenamefont {Ochiai}}]{1742-6596-109-1-012035}%
  \BibitemOpen
  \bibfield  {author} {\bibinfo {author} {\bibfnamefont {Y.}~\bibnamefont
  {Ujiie}}, \bibinfo {author} {\bibfnamefont {N.}~\bibnamefont {Yumoto}},
  \bibinfo {author} {\bibfnamefont {T.}~\bibnamefont {Morimoto}}, \bibinfo
  {author} {\bibfnamefont {N.}~\bibnamefont {Aoki}}, \ and\ \bibinfo {author}
  {\bibfnamefont {Y.}~\bibnamefont {Ochiai}},\ }\href
  {http://stacks.iop.org/1742-6596/109/i=1/a=012035} {\bibfield  {journal}
  {\bibinfo  {journal} {J. Phys. Conf. Ser.}\ }\textbf {\bibinfo {volume}
  {109}},\ \bibinfo {pages} {012035} (\bibinfo {year} {2008})}\BibitemShut
  {NoStop}%
\bibitem [{\citenamefont {Meiss}\ and\ \citenamefont
  {Ott}(1986)}]{MEISS1986387}%
  \BibitemOpen
  \bibfield  {author} {\bibinfo {author} {\bibfnamefont {J.~D.}\ \bibnamefont
  {Meiss}}\ and\ \bibinfo {author} {\bibfnamefont {E.}~\bibnamefont {Ott}},\
  }\href {\doibase http://dx.doi.org/10.1016/0167-2789(86)90041-2} {\bibfield
  {journal} {\bibinfo  {journal} {Physica D}\ }\textbf {\bibinfo {volume}
  {20}},\ \bibinfo {pages} {387 } (\bibinfo {year} {1986})}\BibitemShut
  {NoStop}%
\bibitem [{\citenamefont {Geisel}\ \emph {et~al.}(1987)\citenamefont {Geisel},
  \citenamefont {Zacherl},\ and\ \citenamefont {Radons}}]{PhysRevLett.59.2503}%
  \BibitemOpen
  \bibfield  {author} {\bibinfo {author} {\bibfnamefont {T.}~\bibnamefont
  {Geisel}}, \bibinfo {author} {\bibfnamefont {A.}~\bibnamefont {Zacherl}}, \
  and\ \bibinfo {author} {\bibfnamefont {G.}~\bibnamefont {Radons}},\ }\href
  {\doibase 10.1103/PhysRevLett.59.2503} {\bibfield  {journal} {\bibinfo
  {journal} {Phys. Rev. Lett.}\ }\textbf {\bibinfo {volume} {59}},\ \bibinfo
  {pages} {2503} (\bibinfo {year} {1987})}\BibitemShut {NoStop}%
\bibitem [{\citenamefont {Novoselov}\ \emph {et~al.}(2005)\citenamefont
  {Novoselov}, \citenamefont {Geim}, \citenamefont {Morozov}, \citenamefont
  {Jiang}, \citenamefont {Katsnelson}, \citenamefont {Grigorieva},
  \citenamefont {Dubonos},\ and\ \citenamefont {Firsov}}]{novoselov2005two}%
  \BibitemOpen
  \bibfield  {author} {\bibinfo {author} {\bibfnamefont {K.}~\bibnamefont
  {Novoselov}}, \bibinfo {author} {\bibfnamefont {A.~K.}\ \bibnamefont {Geim}},
  \bibinfo {author} {\bibfnamefont {S.}~\bibnamefont {Morozov}}, \bibinfo
  {author} {\bibfnamefont {D.}~\bibnamefont {Jiang}}, \bibinfo {author}
  {\bibfnamefont {M.}~\bibnamefont {Katsnelson}}, \bibinfo {author}
  {\bibfnamefont {I.}~\bibnamefont {Grigorieva}}, \bibinfo {author}
  {\bibfnamefont {S.}~\bibnamefont {Dubonos}}, \ and\ \bibinfo {author}
  {\bibfnamefont {A.}~\bibnamefont {Firsov}},\ }\href {\doibase
  10.1038/nature04233} {\bibfield  {journal} {\bibinfo  {journal} {Nature}\
  }\textbf {\bibinfo {volume} {438}},\ \bibinfo {pages} {197} (\bibinfo {year}
  {2005})}\BibitemShut {NoStop}%
\bibitem [{\citenamefont {Chen}\ \emph {et~al.}(2010)\citenamefont {Chen},
  \citenamefont {Bae}, \citenamefont {Chialvo}, \citenamefont {Dirks},
  \citenamefont {Bezryadin},\ and\ \citenamefont
  {Mason}}]{0953-8984-22-20-205301}%
  \BibitemOpen
  \bibfield  {author} {\bibinfo {author} {\bibfnamefont {Y.-F.}\ \bibnamefont
  {Chen}}, \bibinfo {author} {\bibfnamefont {M.-H.}\ \bibnamefont {Bae}},
  \bibinfo {author} {\bibfnamefont {C.}~\bibnamefont {Chialvo}}, \bibinfo
  {author} {\bibfnamefont {T.}~\bibnamefont {Dirks}}, \bibinfo {author}
  {\bibfnamefont {A.}~\bibnamefont {Bezryadin}}, \ and\ \bibinfo {author}
  {\bibfnamefont {N.}~\bibnamefont {Mason}},\ }\href
  {http://stacks.iop.org/0953-8984/22/i=20/a=205301} {\bibfield  {journal}
  {\bibinfo  {journal} {Journal of Physics: Condensed Matter}\ }\textbf
  {\bibinfo {volume} {22}},\ \bibinfo {pages} {205301} (\bibinfo {year}
  {2010})}\BibitemShut {NoStop}%
\bibitem [{\citenamefont {Berger}\ \emph {et~al.}(2006)\citenamefont {Berger},
  \citenamefont {Song}, \citenamefont {Li}, \citenamefont {Wu}, \citenamefont
  {Brown}, \citenamefont {Naud}, \citenamefont {Mayou}, \citenamefont {Li},
  \citenamefont {Hass}, \citenamefont {Marchenkov}, \citenamefont {Conrad},
  \citenamefont {First},\ and\ \citenamefont {de~Heer}}]{Berger1191}%
  \BibitemOpen
  \bibfield  {author} {\bibinfo {author} {\bibfnamefont {C.}~\bibnamefont
  {Berger}}, \bibinfo {author} {\bibfnamefont {Z.}~\bibnamefont {Song}},
  \bibinfo {author} {\bibfnamefont {X.}~\bibnamefont {Li}}, \bibinfo {author}
  {\bibfnamefont {X.}~\bibnamefont {Wu}}, \bibinfo {author} {\bibfnamefont
  {N.}~\bibnamefont {Brown}}, \bibinfo {author} {\bibfnamefont
  {C.}~\bibnamefont {Naud}}, \bibinfo {author} {\bibfnamefont {D.}~\bibnamefont
  {Mayou}}, \bibinfo {author} {\bibfnamefont {T.}~\bibnamefont {Li}}, \bibinfo
  {author} {\bibfnamefont {J.}~\bibnamefont {Hass}}, \bibinfo {author}
  {\bibfnamefont {A.~N.}\ \bibnamefont {Marchenkov}}, \bibinfo {author}
  {\bibfnamefont {E.~H.}\ \bibnamefont {Conrad}}, \bibinfo {author}
  {\bibfnamefont {P.~N.}\ \bibnamefont {First}}, \ and\ \bibinfo {author}
  {\bibfnamefont {W.~A.}\ \bibnamefont {de~Heer}},\ }\href {\doibase
  10.1126/science.1125925} {\bibfield  {journal} {\bibinfo  {journal}
  {Science}\ }\textbf {\bibinfo {volume} {312}},\ \bibinfo {pages} {1191}
  (\bibinfo {year} {2006})},\ \Eprint
  {http://arxiv.org/abs/http://science.sciencemag.org/content/312/5777/1191.full.pdf}
  {http://science.sciencemag.org/content/312/5777/1191.full.pdf} \BibitemShut
  {NoStop}%
\bibitem [{\citenamefont {Bohra}\ \emph {et~al.}(2012)\citenamefont {Bohra},
  \citenamefont {Somphonsane}, \citenamefont {Ferry},\ and\ \citenamefont
  {Bird}}]{Bohra2012}%
  \BibitemOpen
  \bibfield  {author} {\bibinfo {author} {\bibfnamefont {G.}~\bibnamefont
  {Bohra}}, \bibinfo {author} {\bibfnamefont {R.}~\bibnamefont {Somphonsane}},
  \bibinfo {author} {\bibfnamefont {D.~K.}\ \bibnamefont {Ferry}}, \ and\
  \bibinfo {author} {\bibfnamefont {J.~P.}\ \bibnamefont {Bird}},\ }\href
  {\doibase 10.1063/1.4748167} {\bibfield  {journal} {\bibinfo  {journal}
  {Applied Physics Letters}\ }\textbf {\bibinfo {volume} {101}},\ \bibinfo
  {pages} {093110} (\bibinfo {year} {2012})},\ \Eprint
  {http://arxiv.org/abs/http://dx.doi.org/10.1063/1.4748167}
  {http://dx.doi.org/10.1063/1.4748167} \BibitemShut {NoStop}%
\bibitem [{\citenamefont {Tikhonenko}\ \emph {et~al.}(2008)\citenamefont
  {Tikhonenko}, \citenamefont {Horsell}, \citenamefont {Gorbachev},\ and\
  \citenamefont {Savchenko}}]{PhysRevLett.100.056802}%
  \BibitemOpen
  \bibfield  {author} {\bibinfo {author} {\bibfnamefont {F.~V.}\ \bibnamefont
  {Tikhonenko}}, \bibinfo {author} {\bibfnamefont {D.~W.}\ \bibnamefont
  {Horsell}}, \bibinfo {author} {\bibfnamefont {R.~V.}\ \bibnamefont
  {Gorbachev}}, \ and\ \bibinfo {author} {\bibfnamefont {A.~K.}\ \bibnamefont
  {Savchenko}},\ }\href {\doibase 10.1103/PhysRevLett.100.056802} {\bibfield
  {journal} {\bibinfo  {journal} {Phys. Rev. Lett.}\ }\textbf {\bibinfo
  {volume} {100}},\ \bibinfo {pages} {056802} (\bibinfo {year}
  {2008})}\BibitemShut {NoStop}%
\bibitem [{\citenamefont {Horsell}\ \emph {et~al.}(2009)\citenamefont
  {Horsell}, \citenamefont {Savchenko}, \citenamefont {Tikhonenko},
  \citenamefont {Kechedzhi}, \citenamefont {Lerner},\ and\ \citenamefont
  {Fal¡¯ko}}]{Horsell20091041}%
  \BibitemOpen
  \bibfield  {author} {\bibinfo {author} {\bibfnamefont {D.}~\bibnamefont
  {Horsell}}, \bibinfo {author} {\bibfnamefont {A.}~\bibnamefont {Savchenko}},
  \bibinfo {author} {\bibfnamefont {F.}~\bibnamefont {Tikhonenko}}, \bibinfo
  {author} {\bibfnamefont {K.}~\bibnamefont {Kechedzhi}}, \bibinfo {author}
  {\bibfnamefont {I.}~\bibnamefont {Lerner}}, \ and\ \bibinfo {author}
  {\bibfnamefont {V.}~\bibnamefont {Fal¡¯ko}},\ }\href {\doibase
  http://dx.doi.org/10.1016/j.ssc.2009.02.058} {\bibfield  {journal} {\bibinfo
  {journal} {Solid State Communications}\ }\textbf {\bibinfo {volume} {149}},\
  \bibinfo {pages} {1041 } (\bibinfo {year} {2009})},\ \bibinfo {note} {recent
  Progress in Graphene Studies}\BibitemShut {NoStop}%
\bibitem [{\citenamefont {Gorbachev}\ \emph {et~al.}(2007)\citenamefont
  {Gorbachev}, \citenamefont {Tikhonenko}, \citenamefont {Mayorov},
  \citenamefont {Horsell},\ and\ \citenamefont
  {Savchenko}}]{PhysRevLett.98.176805}%
  \BibitemOpen
  \bibfield  {author} {\bibinfo {author} {\bibfnamefont {R.~V.}\ \bibnamefont
  {Gorbachev}}, \bibinfo {author} {\bibfnamefont {F.~V.}\ \bibnamefont
  {Tikhonenko}}, \bibinfo {author} {\bibfnamefont {A.~S.}\ \bibnamefont
  {Mayorov}}, \bibinfo {author} {\bibfnamefont {D.~W.}\ \bibnamefont
  {Horsell}}, \ and\ \bibinfo {author} {\bibfnamefont {A.~K.}\ \bibnamefont
  {Savchenko}},\ }\href {\doibase 10.1103/PhysRevLett.98.176805} {\bibfield
  {journal} {\bibinfo  {journal} {Phys. Rev. Lett.}\ }\textbf {\bibinfo
  {volume} {98}},\ \bibinfo {pages} {176805} (\bibinfo {year}
  {2007})}\BibitemShut {NoStop}%
\bibitem [{\citenamefont {Mohanty}\ and\ \citenamefont
  {Webb}(2003)}]{PhysRevLett.91.066604}%
  \BibitemOpen
  \bibfield  {author} {\bibinfo {author} {\bibfnamefont {P.}~\bibnamefont
  {Mohanty}}\ and\ \bibinfo {author} {\bibfnamefont {R.~A.}\ \bibnamefont
  {Webb}},\ }\href {\doibase 10.1103/PhysRevLett.91.066604} {\bibfield
  {journal} {\bibinfo  {journal} {Phys. Rev. Lett.}\ }\textbf {\bibinfo
  {volume} {91}},\ \bibinfo {pages} {066604} (\bibinfo {year}
  {2003})}\BibitemShut {NoStop}%
\bibitem [{\citenamefont {Mohanty}\ and\ \citenamefont
  {Webb}(1997)}]{PhysRevB.55.R13452}%
  \BibitemOpen
  \bibfield  {author} {\bibinfo {author} {\bibfnamefont {P.}~\bibnamefont
  {Mohanty}}\ and\ \bibinfo {author} {\bibfnamefont {R.~A.}\ \bibnamefont
  {Webb}},\ }\href {\doibase 10.1103/PhysRevB.55.R13452} {\bibfield  {journal}
  {\bibinfo  {journal} {Phys. Rev. B}\ }\textbf {\bibinfo {volume} {55}},\
  \bibinfo {pages} {R13452} (\bibinfo {year} {1997})}\BibitemShut {NoStop}%
\bibitem [{\citenamefont {Pierre}\ \emph {et~al.}(2003)\citenamefont {Pierre},
  \citenamefont {Gougam}, \citenamefont {Anthore}, \citenamefont {Pothier},
  \citenamefont {Esteve},\ and\ \citenamefont {Birge}}]{PhysRevB.68.085413}%
  \BibitemOpen
  \bibfield  {author} {\bibinfo {author} {\bibfnamefont {F.}~\bibnamefont
  {Pierre}}, \bibinfo {author} {\bibfnamefont {A.~B.}\ \bibnamefont {Gougam}},
  \bibinfo {author} {\bibfnamefont {A.}~\bibnamefont {Anthore}}, \bibinfo
  {author} {\bibfnamefont {H.}~\bibnamefont {Pothier}}, \bibinfo {author}
  {\bibfnamefont {D.}~\bibnamefont {Esteve}}, \ and\ \bibinfo {author}
  {\bibfnamefont {N.~O.}\ \bibnamefont {Birge}},\ }\href {\doibase
  10.1103/PhysRevB.68.085413} {\bibfield  {journal} {\bibinfo  {journal} {Phys.
  Rev. B}\ }\textbf {\bibinfo {volume} {68}},\ \bibinfo {pages} {085413}
  (\bibinfo {year} {2003})}\BibitemShut {NoStop}%
\bibitem [{\citenamefont {Liu}\ \emph {et~al.}(2016)\citenamefont {Liu},
  \citenamefont {Akis}, \citenamefont {Ferry}, \citenamefont {Bohra},
  \citenamefont {Somphonsane}, \citenamefont {Ramamoorthy},\ and\ \citenamefont
  {Bird}}]{Liu2016}%
  \BibitemOpen
  \bibfield  {author} {\bibinfo {author} {\bibfnamefont {B.}~\bibnamefont
  {Liu}}, \bibinfo {author} {\bibfnamefont {R.}~\bibnamefont {Akis}}, \bibinfo
  {author} {\bibfnamefont {D.~K.}\ \bibnamefont {Ferry}}, \bibinfo {author}
  {\bibfnamefont {G.}~\bibnamefont {Bohra}}, \bibinfo {author} {\bibfnamefont
  {R.}~\bibnamefont {Somphonsane}}, \bibinfo {author} {\bibfnamefont
  {H.}~\bibnamefont {Ramamoorthy}}, \ and\ \bibinfo {author} {\bibfnamefont
  {J.~P.}\ \bibnamefont {Bird}},\ }\href
  {http://stacks.iop.org/0953-8984/28/i=13/a=135302} {\bibfield  {journal}
  {\bibinfo  {journal} {Journal of Physics: Condensed Matter}\ }\textbf
  {\bibinfo {volume} {28}},\ \bibinfo {pages} {135302} (\bibinfo {year}
  {2016})}\BibitemShut {NoStop}%
\bibitem [{\citenamefont {Gu}\ and\ \citenamefont
  {Zhou}(2010)}]{PhysRevE.82.011136}%
  \BibitemOpen
  \bibfield  {author} {\bibinfo {author} {\bibfnamefont {G.-F.}\ \bibnamefont
  {Gu}}\ and\ \bibinfo {author} {\bibfnamefont {W.-X.}\ \bibnamefont {Zhou}},\
  }\href {\doibase 10.1103/PhysRevE.82.011136} {\bibfield  {journal} {\bibinfo
  {journal} {Phys. Rev. E}\ }\textbf {\bibinfo {volume} {82}},\ \bibinfo
  {pages} {011136} (\bibinfo {year} {2010})}\BibitemShut {NoStop}%
\bibitem [{\citenamefont {Zhou}(2009)}]{Zhou2009}%
  \BibitemOpen
  \bibfield  {author} {\bibinfo {author} {\bibfnamefont {W.-X.}\ \bibnamefont
  {Zhou}},\ }\href {http://stacks.iop.org/0295-5075/88/i=2/a=28004} {\bibfield
  {journal} {\bibinfo  {journal} {EPL (Europhysics Letters)}\ }\textbf
  {\bibinfo {volume} {88}},\ \bibinfo {pages} {28004} (\bibinfo {year}
  {2009})}\BibitemShut {NoStop}%
\bibitem [{\citenamefont {Grassberger}\ and\ \citenamefont
  {Procaccia}(1983)}]{PhysRevLett.50.346}%
  \BibitemOpen
  \bibfield  {author} {\bibinfo {author} {\bibfnamefont {P.}~\bibnamefont
  {Grassberger}}\ and\ \bibinfo {author} {\bibfnamefont {I.}~\bibnamefont
  {Procaccia}},\ }\href {\doibase 10.1103/PhysRevLett.50.346} {\bibfield
  {journal} {\bibinfo  {journal} {Phys. Rev. Lett.}\ }\textbf {\bibinfo
  {volume} {50}},\ \bibinfo {pages} {346} (\bibinfo {year} {1983})}\BibitemShut
  {NoStop}%
\bibitem [{\citenamefont {Kellay}\ and\ \citenamefont
  {Goldburg}(2002)}]{Kellay2002}%
  \BibitemOpen
  \bibfield  {author} {\bibinfo {author} {\bibfnamefont {H.}~\bibnamefont
  {Kellay}}\ and\ \bibinfo {author} {\bibfnamefont {W.~I.}\ \bibnamefont
  {Goldburg}},\ }\href {http://stacks.iop.org/0034-4885/65/i=5/a=204}
  {\bibfield  {journal} {\bibinfo  {journal} {Reports on Progress in Physics}\
  }\textbf {\bibinfo {volume} {65}},\ \bibinfo {pages} {845} (\bibinfo {year}
  {2002})}\BibitemShut {NoStop}%
\bibitem [{\citenamefont {Boffetta}\ and\ \citenamefont
  {Ecke}(2012)}]{Boffetta2012}%
  \BibitemOpen
  \bibfield  {author} {\bibinfo {author} {\bibfnamefont {G.}~\bibnamefont
  {Boffetta}}\ and\ \bibinfo {author} {\bibfnamefont {R.~E.}\ \bibnamefont
  {Ecke}},\ }\href {\doibase 10.1146/annurev-fluid-120710-101240} {\bibfield
  {journal} {\bibinfo  {journal} {Annual Review of Fluid Mechanics}\ }\textbf
  {\bibinfo {volume} {44}},\ \bibinfo {pages} {427} (\bibinfo {year} {2012})},\
  \Eprint
  {http://arxiv.org/abs/https://doi.org/10.1146/annurev-fluid-120710-101240}
  {https://doi.org/10.1146/annurev-fluid-120710-101240} \BibitemShut {NoStop}%
\bibitem [{\citenamefont {Heller}(1984)}]{PhysRevLett.53.1515}%
  \BibitemOpen
  \bibfield  {author} {\bibinfo {author} {\bibfnamefont {E.~J.}\ \bibnamefont
  {Heller}},\ }\href {\doibase 10.1103/PhysRevLett.53.1515} {\bibfield
  {journal} {\bibinfo  {journal} {Phys. Rev. Lett.}\ }\textbf {\bibinfo
  {volume} {53}},\ \bibinfo {pages} {1515} (\bibinfo {year}
  {1984})}\BibitemShut {NoStop}%
\bibitem [{\citenamefont {Harper}(1955)}]{0370-1298-68-10-305}%
  \BibitemOpen
  \bibfield  {author} {\bibinfo {author} {\bibfnamefont {P.~G.}\ \bibnamefont
  {Harper}},\ }\href {http://stacks.iop.org/0370-1298/68/i=10/a=305} {\bibfield
   {journal} {\bibinfo  {journal} {Proc. Phys. Soc. London, Sect. A}\ }\textbf
  {\bibinfo {volume} {68}},\ \bibinfo {pages} {879} (\bibinfo {year}
  {1955})}\BibitemShut {NoStop}%
\bibitem [{\citenamefont {Hofstadter}(1976)}]{PhysRevB.14.2239}%
  \BibitemOpen
  \bibfield  {author} {\bibinfo {author} {\bibfnamefont {D.~R.}\ \bibnamefont
  {Hofstadter}},\ }\href {\doibase 10.1103/PhysRevB.14.2239} {\bibfield
  {journal} {\bibinfo  {journal} {Phys. Rev. B}\ }\textbf {\bibinfo {volume}
  {14}},\ \bibinfo {pages} {2239} (\bibinfo {year} {1976})}\BibitemShut
  {NoStop}%
\bibitem [{\citenamefont {Ostlund}\ \emph {et~al.}(1983)\citenamefont
  {Ostlund}, \citenamefont {Pandit}, \citenamefont {Rand}, \citenamefont
  {Schellnhuber},\ and\ \citenamefont {Siggia}}]{PhysRevLett.50.1873}%
  \BibitemOpen
  \bibfield  {author} {\bibinfo {author} {\bibfnamefont {S.}~\bibnamefont
  {Ostlund}}, \bibinfo {author} {\bibfnamefont {R.}~\bibnamefont {Pandit}},
  \bibinfo {author} {\bibfnamefont {D.}~\bibnamefont {Rand}}, \bibinfo {author}
  {\bibfnamefont {H.~J.}\ \bibnamefont {Schellnhuber}}, \ and\ \bibinfo
  {author} {\bibfnamefont {E.~D.}\ \bibnamefont {Siggia}},\ }\href {\doibase
  10.1103/PhysRevLett.50.1873} {\bibfield  {journal} {\bibinfo  {journal}
  {Phys. Rev. Lett.}\ }\textbf {\bibinfo {volume} {50}},\ \bibinfo {pages}
  {1873} (\bibinfo {year} {1983})}\BibitemShut {NoStop}%
\bibitem [{\citenamefont {Sena}\ \emph {et~al.}(2010)\citenamefont {Sena},
  \citenamefont {Jr}, \citenamefont {Farias}, \citenamefont {Vasconcelos},\
  and\ \citenamefont {Albuquerque}}]{0953-8984-22-46-465305}%
  \BibitemOpen
  \bibfield  {author} {\bibinfo {author} {\bibfnamefont {S.~H.~R.}\
  \bibnamefont {Sena}}, \bibinfo {author} {\bibfnamefont {J.~M.~P.}\
  \bibnamefont {Jr}}, \bibinfo {author} {\bibfnamefont {G.~A.}\ \bibnamefont
  {Farias}}, \bibinfo {author} {\bibfnamefont {M.~S.}\ \bibnamefont
  {Vasconcelos}}, \ and\ \bibinfo {author} {\bibfnamefont {E.~L.}\ \bibnamefont
  {Albuquerque}},\ }\href {http://stacks.iop.org/0953-8984/22/i=46/a=465305}
  {\bibfield  {journal} {\bibinfo  {journal} {J. Phys.: Condens. Matter}\
  }\textbf {\bibinfo {volume} {22}},\ \bibinfo {pages} {465305} (\bibinfo
  {year} {2010})}\BibitemShut {NoStop}%
\bibitem [{\citenamefont {Zurek}(1981)}]{PhysRevD.24.1516}%
  \BibitemOpen
  \bibfield  {author} {\bibinfo {author} {\bibfnamefont {W.~H.}\ \bibnamefont
  {Zurek}},\ }\href {\doibase 10.1103/PhysRevD.24.1516} {\bibfield  {journal}
  {\bibinfo  {journal} {Phys. Rev. D}\ }\textbf {\bibinfo {volume} {24}},\
  \bibinfo {pages} {1516} (\bibinfo {year} {1981})}\BibitemShut {NoStop}%
\bibitem [{\citenamefont {Huang}\ \emph {et~al.}(2009)\citenamefont {Huang},
  \citenamefont {Lai}, \citenamefont {Ferry}, \citenamefont {Goodnick},\ and\
  \citenamefont {Akis}}]{PhysRevLett.103.054101}%
  \BibitemOpen
  \bibfield  {author} {\bibinfo {author} {\bibfnamefont {L.}~\bibnamefont
  {Huang}}, \bibinfo {author} {\bibfnamefont {Y.-C.}\ \bibnamefont {Lai}},
  \bibinfo {author} {\bibfnamefont {D.~K.}\ \bibnamefont {Ferry}}, \bibinfo
  {author} {\bibfnamefont {S.~M.}\ \bibnamefont {Goodnick}}, \ and\ \bibinfo
  {author} {\bibfnamefont {R.}~\bibnamefont {Akis}},\ }\href {\doibase
  10.1103/PhysRevLett.103.054101} {\bibfield  {journal} {\bibinfo  {journal}
  {Phys. Rev. Lett.}\ }\textbf {\bibinfo {volume} {103}},\ \bibinfo {pages}
  {054101} (\bibinfo {year} {2009})}\BibitemShut {NoStop}%
\bibitem [{\citenamefont {Ostlund}\ and\ \citenamefont
  {Pandit}(1984)}]{PhysRevB.29.1394}%
  \BibitemOpen
  \bibfield  {author} {\bibinfo {author} {\bibfnamefont {S.}~\bibnamefont
  {Ostlund}}\ and\ \bibinfo {author} {\bibfnamefont {R.}~\bibnamefont
  {Pandit}},\ }\href {\doibase 10.1103/PhysRevB.29.1394} {\bibfield  {journal}
  {\bibinfo  {journal} {Phys. Rev. B}\ }\textbf {\bibinfo {volume} {29}},\
  \bibinfo {pages} {1394} (\bibinfo {year} {1984})}\BibitemShut {NoStop}%
\bibitem [{\citenamefont {Brandes}\ \emph {et~al.}(1996)\citenamefont
  {Brandes}, \citenamefont {Huckestein},\ and\ \citenamefont
  {Schweitzer}}]{ANDP:ANDP2065080803}%
  \BibitemOpen
  \bibfield  {author} {\bibinfo {author} {\bibfnamefont {T.}~\bibnamefont
  {Brandes}}, \bibinfo {author} {\bibfnamefont {B.}~\bibnamefont {Huckestein}},
  \ and\ \bibinfo {author} {\bibfnamefont {L.}~\bibnamefont {Schweitzer}},\
  }\href {\doibase 10.1002/andp.2065080803} {\bibfield  {journal} {\bibinfo
  {journal} {Ann. Phys.}\ }\textbf {\bibinfo {volume} {508}},\ \bibinfo {pages}
  {633} (\bibinfo {year} {1996})}\BibitemShut {NoStop}%
\bibitem [{\citenamefont {Harris}\ and\ \citenamefont
  {Aharony}(1987)}]{0295-5075-4-12-003}%
  \BibitemOpen
  \bibfield  {author} {\bibinfo {author} {\bibfnamefont {A.~B.}\ \bibnamefont
  {Harris}}\ and\ \bibinfo {author} {\bibfnamefont {A.}~\bibnamefont
  {Aharony}},\ }\href {http://stacks.iop.org/0295-5075/4/i=12/a=003} {\bibfield
   {journal} {\bibinfo  {journal} {Europhys. Lett.}\ }\textbf {\bibinfo
  {volume} {4}},\ \bibinfo {pages} {1355} (\bibinfo {year} {1987})}\BibitemShut
  {NoStop}%
\bibitem [{\citenamefont {Benenti}\ \emph {et~al.}(2001)\citenamefont
  {Benenti}, \citenamefont {Casati}, \citenamefont {Guarneri},\ and\
  \citenamefont {Terraneo}}]{PhysRevLett.87.014101}%
  \BibitemOpen
  \bibfield  {author} {\bibinfo {author} {\bibfnamefont {G.}~\bibnamefont
  {Benenti}}, \bibinfo {author} {\bibfnamefont {G.}~\bibnamefont {Casati}},
  \bibinfo {author} {\bibfnamefont {I.}~\bibnamefont {Guarneri}}, \ and\
  \bibinfo {author} {\bibfnamefont {M.}~\bibnamefont {Terraneo}},\ }\href
  {\doibase 10.1103/PhysRevLett.87.014101} {\bibfield  {journal} {\bibinfo
  {journal} {Phys. Rev. Lett.}\ }\textbf {\bibinfo {volume} {87}},\ \bibinfo
  {pages} {014101} (\bibinfo {year} {2001})}\BibitemShut {NoStop}%
\bibitem [{\citenamefont {Ugeda}\ \emph {et~al.}(2010)\citenamefont {Ugeda},
  \citenamefont {Brihuega}, \citenamefont {Guinea},\ and\ \citenamefont
  {G\'omez-Rodr\'{\i}guez}}]{PhysRevLett.104.096804}%
  \BibitemOpen
  \bibfield  {author} {\bibinfo {author} {\bibfnamefont {M.~M.}\ \bibnamefont
  {Ugeda}}, \bibinfo {author} {\bibfnamefont {I.}~\bibnamefont {Brihuega}},
  \bibinfo {author} {\bibfnamefont {F.}~\bibnamefont {Guinea}}, \ and\ \bibinfo
  {author} {\bibfnamefont {J.~M.}\ \bibnamefont {G\'omez-Rodr\'{\i}guez}},\
  }\href {\doibase 10.1103/PhysRevLett.104.096804} {\bibfield  {journal}
  {\bibinfo  {journal} {Phys. Rev. Lett.}\ }\textbf {\bibinfo {volume} {104}},\
  \bibinfo {pages} {096804} (\bibinfo {year} {2010})}\BibitemShut {NoStop}%
\bibitem [{\citenamefont {Shytov}\ \emph {et~al.}(2007)\citenamefont {Shytov},
  \citenamefont {Katsnelson},\ and\ \citenamefont
  {Levitov}}]{PhysRevLett.99.236801}%
  \BibitemOpen
  \bibfield  {author} {\bibinfo {author} {\bibfnamefont {A.~V.}\ \bibnamefont
  {Shytov}}, \bibinfo {author} {\bibfnamefont {M.~I.}\ \bibnamefont
  {Katsnelson}}, \ and\ \bibinfo {author} {\bibfnamefont {L.~S.}\ \bibnamefont
  {Levitov}},\ }\href {\doibase 10.1103/PhysRevLett.99.236801} {\bibfield
  {journal} {\bibinfo  {journal} {Phys. Rev. Lett.}\ }\textbf {\bibinfo
  {volume} {99}},\ \bibinfo {pages} {236801} (\bibinfo {year}
  {2007})}\BibitemShut {NoStop}%
\bibitem [{\citenamefont {Das~Sarma}\ \emph {et~al.}(2011)\citenamefont
  {Das~Sarma}, \citenamefont {Adam}, \citenamefont {Hwang},\ and\ \citenamefont
  {Rossi}}]{RevModPhys.83.407}%
  \BibitemOpen
  \bibfield  {author} {\bibinfo {author} {\bibfnamefont {S.}~\bibnamefont
  {Das~Sarma}}, \bibinfo {author} {\bibfnamefont {S.}~\bibnamefont {Adam}},
  \bibinfo {author} {\bibfnamefont {E.~H.}\ \bibnamefont {Hwang}}, \ and\
  \bibinfo {author} {\bibfnamefont {E.}~\bibnamefont {Rossi}},\ }\href
  {\doibase 10.1103/RevModPhys.83.407} {\bibfield  {journal} {\bibinfo
  {journal} {Rev. Mod. Phys.}\ }\textbf {\bibinfo {volume} {83}},\ \bibinfo
  {pages} {407} (\bibinfo {year} {2011})}\BibitemShut {NoStop}%
\bibitem [{\citenamefont {Casati}\ \emph {et~al.}(2000)\citenamefont {Casati},
  \citenamefont {Guarneri},\ and\ \citenamefont {Maspero}}]{PhysRevLett.84.63}%
  \BibitemOpen
  \bibfield  {author} {\bibinfo {author} {\bibfnamefont {G.}~\bibnamefont
  {Casati}}, \bibinfo {author} {\bibfnamefont {I.}~\bibnamefont {Guarneri}}, \
  and\ \bibinfo {author} {\bibfnamefont {G.}~\bibnamefont {Maspero}},\ }\href
  {\doibase 10.1103/PhysRevLett.84.63} {\bibfield  {journal} {\bibinfo
  {journal} {Phys. Rev. Lett.}\ }\textbf {\bibinfo {volume} {84}},\ \bibinfo
  {pages} {63} (\bibinfo {year} {2000})}\BibitemShut {NoStop}%
\bibitem [{\citenamefont {Lai}\ \emph {et~al.}(1992)\citenamefont {Lai},
  \citenamefont {Bl\"umel}, \citenamefont {Ott},\ and\ \citenamefont
  {Grebogi}}]{PhysRevLett.68.3491}%
  \BibitemOpen
  \bibfield  {author} {\bibinfo {author} {\bibfnamefont {Y.-C.}\ \bibnamefont
  {Lai}}, \bibinfo {author} {\bibfnamefont {R.}~\bibnamefont {Bl\"umel}},
  \bibinfo {author} {\bibfnamefont {E.}~\bibnamefont {Ott}}, \ and\ \bibinfo
  {author} {\bibfnamefont {C.}~\bibnamefont {Grebogi}},\ }\href {\doibase
  10.1103/PhysRevLett.68.3491} {\bibfield  {journal} {\bibinfo  {journal}
  {Phys. Rev. Lett.}\ }\textbf {\bibinfo {volume} {68}},\ \bibinfo {pages}
  {3491} (\bibinfo {year} {1992})}\BibitemShut {NoStop}%
\bibitem [{\citenamefont {Nakayama}\ and\ \citenamefont
  {Yakubo}(2013)}]{nakayama2013fractal}%
  \BibitemOpen
  \bibfield  {author} {\bibinfo {author} {\bibfnamefont {T.}~\bibnamefont
  {Nakayama}}\ and\ \bibinfo {author} {\bibfnamefont {K.}~\bibnamefont
  {Yakubo}},\ }\href {https://books.google.co.in/books?id=D2r7CAAAQBAJ} {\emph
  {\bibinfo {title} {Fractal Concepts in Condensed Matter Physics}}},\ Springer
  Series in Solid-State Sciences\ (\bibinfo  {publisher} {Springer Berlin
  Heidelberg},\ \bibinfo {year} {2013})\BibitemShut {NoStop}%
\end{thebibliography}

\begin{thebibliography}{16}%
\makeatletter
\providecommand \@ifxundefined [1]{%
 \@ifx{#1\undefined}
}%
\providecommand \@ifnum [1]{%
 \ifnum #1\expandafter \@firstoftwo
 \else \expandafter \@secondoftwo
 \fi
}%
\providecommand \@ifx [1]{%
 \ifx #1\expandafter \@firstoftwo
 \else \expandafter \@secondoftwo
 \fi
}%
\providecommand \natexlab [1]{#1}%
\providecommand \enquote  [1]{``#1''}%
\providecommand \bibnamefont  [1]{#1}%
\providecommand \bibfnamefont [1]{#1}%
\providecommand \citenamefont [1]{#1}%
\providecommand \href@noop [0]{\@secondoftwo}%
\providecommand \href [0]{\begingroup \@sanitize@url \@href}%
\providecommand \@href[1]{\@@startlink{#1}\@@href}%
\providecommand \@@href[1]{\endgroup#1\@@endlink}%
\providecommand \@sanitize@url [0]{\catcode `\\12\catcode `\$12\catcode
  `\&12\catcode `\#12\catcode `\^12\catcode `\_12\catcode `\%12\relax}%
\providecommand \@@startlink[1]{}%
\providecommand \@@endlink[0]{}%
\providecommand \url  [0]{\begingroup\@sanitize@url \@url }%
\providecommand \@url [1]{\endgroup\@href {#1}{\urlprefix }}%
\providecommand \urlprefix  [0]{URL }%
\providecommand \Eprint [0]{\href }%
\providecommand \doibase [0]{http://dx.doi.org/}%
\providecommand \selectlanguage [0]{\@gobble}%
\providecommand \bibinfo  [0]{\@secondoftwo}%
\providecommand \bibfield  [0]{\@secondoftwo}%
\providecommand \translation [1]{[#1]}%
\providecommand \BibitemOpen [0]{}%
\providecommand \bibitemStop [0]{}%
\providecommand \bibitemNoStop [0]{.\EOS\space}%
\providecommand \EOS [0]{\spacefactor3000\relax}%
\providecommand \BibitemShut  [1]{\csname bibitem#1\endcsname}%
\let\auto@bib@innerbib\@empty
\bibitem [{\citenamefont {Beenakker}\ and\ \citenamefont {van
  Houten}(1991)}]{SBeenakker19911}%
  \BibitemOpen
  \bibfield  {author} {\bibinfo {author} {\bibfnamefont {C.}~\bibnamefont
  {Beenakker}}\ and\ \bibinfo {author} {\bibfnamefont {H.}~\bibnamefont {van
  Houten}},\ }in\ \href {\doibase
  http://dx.doi.org/10.1016/S0081-1947(08)60091-0} {\emph {\bibinfo {booktitle}
  {Semiconductor Heterostructures and Nanostructures}}},\ \bibinfo {series}
  {Solid State Physics}, Vol.~\bibinfo {volume} {44},\ \bibinfo {editor}
  {edited by\ \bibinfo {editor} {\bibfnamefont {H.}~\bibnamefont {Ehrenreich}}\
  and\ \bibinfo {editor} {\bibfnamefont {D.}~\bibnamefont {Turnbull}}}\
  (\bibinfo  {publisher} {Academic Press},\ \bibinfo {year} {1991})\ pp.\
  \bibinfo {pages} {1 -- 228}\BibitemShut {NoStop}%
\bibitem [{\citenamefont {Tikhonenko}\ \emph {et~al.}(2008)\citenamefont
  {Tikhonenko}, \citenamefont {Horsell}, \citenamefont {Gorbachev},\ and\
  \citenamefont {Savchenko}}]{SPhysRevLett.100.056802}%
  \BibitemOpen
  \bibfield  {author} {\bibinfo {author} {\bibfnamefont {F.~V.}\ \bibnamefont
  {Tikhonenko}}, \bibinfo {author} {\bibfnamefont {D.~W.}\ \bibnamefont
  {Horsell}}, \bibinfo {author} {\bibfnamefont {R.~V.}\ \bibnamefont
  {Gorbachev}}, \ and\ \bibinfo {author} {\bibfnamefont {A.~K.}\ \bibnamefont
  {Savchenko}},\ }\href {\doibase 10.1103/PhysRevLett.100.056802} {\bibfield
  {journal} {\bibinfo  {journal} {Phys. Rev. Lett.}\ }\textbf {\bibinfo
  {volume} {100}},\ \bibinfo {pages} {056802} (\bibinfo {year}
  {2008})}\BibitemShut {NoStop}%
\bibitem [{\citenamefont {Horsell}\ \emph {et~al.}(2009)\citenamefont
  {Horsell}, \citenamefont {Savchenko}, \citenamefont {Tikhonenko},
  \citenamefont {Kechedzhi}, \citenamefont {Lerner},\ and\ \citenamefont
  {Fal¡¯ko}}]{SHorsell20091041}%
  \BibitemOpen
  \bibfield  {author} {\bibinfo {author} {\bibfnamefont {D.}~\bibnamefont
  {Horsell}}, \bibinfo {author} {\bibfnamefont {A.}~\bibnamefont {Savchenko}},
  \bibinfo {author} {\bibfnamefont {F.}~\bibnamefont {Tikhonenko}}, \bibinfo
  {author} {\bibfnamefont {K.}~\bibnamefont {Kechedzhi}}, \bibinfo {author}
  {\bibfnamefont {I.}~\bibnamefont {Lerner}}, \ and\ \bibinfo {author}
  {\bibfnamefont {V.}~\bibnamefont {Fal¡¯ko}},\ }\href {\doibase
  http://dx.doi.org/10.1016/j.ssc.2009.02.058} {\bibfield  {journal} {\bibinfo
  {journal} {Solid State Communications}\ }\textbf {\bibinfo {volume} {149}},\
  \bibinfo {pages} {1041 } (\bibinfo {year} {2009})},\ \bibinfo {note} {recent
  Progress in Graphene Studies}\BibitemShut {NoStop}%
\bibitem [{\citenamefont {Chen}\ \emph {et~al.}(2010)\citenamefont {Chen},
  \citenamefont {Bae}, \citenamefont {Chialvo}, \citenamefont {Dirks},
  \citenamefont {Bezryadin},\ and\ \citenamefont
  {Mason}}]{S0953-8984-22-20-205301}%
  \BibitemOpen
  \bibfield  {author} {\bibinfo {author} {\bibfnamefont {Y.-F.}\ \bibnamefont
  {Chen}}, \bibinfo {author} {\bibfnamefont {M.-H.}\ \bibnamefont {Bae}},
  \bibinfo {author} {\bibfnamefont {C.}~\bibnamefont {Chialvo}}, \bibinfo
  {author} {\bibfnamefont {T.}~\bibnamefont {Dirks}}, \bibinfo {author}
  {\bibfnamefont {A.}~\bibnamefont {Bezryadin}}, \ and\ \bibinfo {author}
  {\bibfnamefont {N.}~\bibnamefont {Mason}},\ }\href
  {http://stacks.iop.org/0953-8984/22/i=20/a=205301} {\bibfield  {journal}
  {\bibinfo  {journal} {Journal of Physics: Condensed Matter}\ }\textbf
  {\bibinfo {volume} {22}},\ \bibinfo {pages} {205301} (\bibinfo {year}
  {2010})}\BibitemShut {NoStop}%
\bibitem [{\citenamefont {Berger}\ \emph {et~al.}(2006)\citenamefont {Berger},
  \citenamefont {Song}, \citenamefont {Li}, \citenamefont {Wu}, \citenamefont
  {Brown}, \citenamefont {Naud}, \citenamefont {Mayou}, \citenamefont {Li},
  \citenamefont {Hass}, \citenamefont {Marchenkov}, \citenamefont {Conrad},
  \citenamefont {First},\ and\ \citenamefont {de~Heer}}]{SBerger1191}%
  \BibitemOpen
  \bibfield  {author} {\bibinfo {author} {\bibfnamefont {C.}~\bibnamefont
  {Berger}}, \bibinfo {author} {\bibfnamefont {Z.}~\bibnamefont {Song}},
  \bibinfo {author} {\bibfnamefont {X.}~\bibnamefont {Li}}, \bibinfo {author}
  {\bibfnamefont {X.}~\bibnamefont {Wu}}, \bibinfo {author} {\bibfnamefont
  {N.}~\bibnamefont {Brown}}, \bibinfo {author} {\bibfnamefont
  {C.}~\bibnamefont {Naud}}, \bibinfo {author} {\bibfnamefont {D.}~\bibnamefont
  {Mayou}}, \bibinfo {author} {\bibfnamefont {T.}~\bibnamefont {Li}}, \bibinfo
  {author} {\bibfnamefont {J.}~\bibnamefont {Hass}}, \bibinfo {author}
  {\bibfnamefont {A.~N.}\ \bibnamefont {Marchenkov}}, \bibinfo {author}
  {\bibfnamefont {E.~H.}\ \bibnamefont {Conrad}}, \bibinfo {author}
  {\bibfnamefont {P.~N.}\ \bibnamefont {First}}, \ and\ \bibinfo {author}
  {\bibfnamefont {W.~A.}\ \bibnamefont {de~Heer}},\ }\href {\doibase
  10.1126/science.1125925} {\bibfield  {journal} {\bibinfo  {journal}
  {Science}\ }\textbf {\bibinfo {volume} {312}},\ \bibinfo {pages} {1191}
  (\bibinfo {year} {2006})},\ \Eprint
  {http://arxiv.org/abs/http://science.sciencemag.org/content/312/5777/1191.full.pdf}
  {http://science.sciencemag.org/content/312/5777/1191.full.pdf} \BibitemShut
  {NoStop}%
\bibitem [{\citenamefont {Bohra}\ \emph {et~al.}(2012)\citenamefont {Bohra},
  \citenamefont {Somphonsane}, \citenamefont {Ferry},\ and\ \citenamefont
  {Bird}}]{SBohra2012}%
  \BibitemOpen
  \bibfield  {author} {\bibinfo {author} {\bibfnamefont {G.}~\bibnamefont
  {Bohra}}, \bibinfo {author} {\bibfnamefont {R.}~\bibnamefont {Somphonsane}},
  \bibinfo {author} {\bibfnamefont {D.~K.}\ \bibnamefont {Ferry}}, \ and\
  \bibinfo {author} {\bibfnamefont {J.~P.}\ \bibnamefont {Bird}},\ }\href
  {\doibase 10.1063/1.4748167} {\bibfield  {journal} {\bibinfo  {journal}
  {Applied Physics Letters}\ }\textbf {\bibinfo {volume} {101}},\ \bibinfo
  {pages} {093110} (\bibinfo {year} {2012})},\ \Eprint
  {http://arxiv.org/abs/http://dx.doi.org/10.1063/1.4748167}
  {http://dx.doi.org/10.1063/1.4748167} \BibitemShut {NoStop}%
\bibitem [{\citenamefont {Gorbachev}\ \emph {et~al.}(2007)\citenamefont
  {Gorbachev}, \citenamefont {Tikhonenko}, \citenamefont {Mayorov},
  \citenamefont {Horsell},\ and\ \citenamefont
  {Savchenko}}]{SPhysRevLett.98.176805}%
  \BibitemOpen
  \bibfield  {author} {\bibinfo {author} {\bibfnamefont {R.~V.}\ \bibnamefont
  {Gorbachev}}, \bibinfo {author} {\bibfnamefont {F.~V.}\ \bibnamefont
  {Tikhonenko}}, \bibinfo {author} {\bibfnamefont {A.~S.}\ \bibnamefont
  {Mayorov}}, \bibinfo {author} {\bibfnamefont {D.~W.}\ \bibnamefont
  {Horsell}}, \ and\ \bibinfo {author} {\bibfnamefont {A.~K.}\ \bibnamefont
  {Savchenko}},\ }\href {\doibase 10.1103/PhysRevLett.98.176805} {\bibfield
  {journal} {\bibinfo  {journal} {Phys. Rev. Lett.}\ }\textbf {\bibinfo
  {volume} {98}},\ \bibinfo {pages} {176805} (\bibinfo {year}
  {2007})}\BibitemShut {NoStop}%
\bibitem [{\citenamefont {Mohanty}\ and\ \citenamefont
  {Webb}(2003)}]{SPhysRevLett.91.066604}%
  \BibitemOpen
  \bibfield  {author} {\bibinfo {author} {\bibfnamefont {P.}~\bibnamefont
  {Mohanty}}\ and\ \bibinfo {author} {\bibfnamefont {R.~A.}\ \bibnamefont
  {Webb}},\ }\href {\doibase 10.1103/PhysRevLett.91.066604} {\bibfield
  {journal} {\bibinfo  {journal} {Phys. Rev. Lett.}\ }\textbf {\bibinfo
  {volume} {91}},\ \bibinfo {pages} {066604} (\bibinfo {year}
  {2003})}\BibitemShut {NoStop}%
\bibitem [{\citenamefont {Mohanty}\ and\ \citenamefont
  {Webb}(1997)}]{SPhysRevB.55.R13452}%
  \BibitemOpen
  \bibfield  {author} {\bibinfo {author} {\bibfnamefont {P.}~\bibnamefont
  {Mohanty}}\ and\ \bibinfo {author} {\bibfnamefont {R.~A.}\ \bibnamefont
  {Webb}},\ }\href {\doibase 10.1103/PhysRevB.55.R13452} {\bibfield  {journal}
  {\bibinfo  {journal} {Phys. Rev. B}\ }\textbf {\bibinfo {volume} {55}},\
  \bibinfo {pages} {R13452} (\bibinfo {year} {1997})}\BibitemShut {NoStop}%
\bibitem [{\citenamefont {Pierre}\ \emph {et~al.}(2003)\citenamefont {Pierre},
  \citenamefont {Gougam}, \citenamefont {Anthore}, \citenamefont {Pothier},
  \citenamefont {Esteve},\ and\ \citenamefont {Birge}}]{SPhysRevB.68.085413}%
  \BibitemOpen
  \bibfield  {author} {\bibinfo {author} {\bibfnamefont {F.}~\bibnamefont
  {Pierre}}, \bibinfo {author} {\bibfnamefont {A.~B.}\ \bibnamefont {Gougam}},
  \bibinfo {author} {\bibfnamefont {A.}~\bibnamefont {Anthore}}, \bibinfo
  {author} {\bibfnamefont {H.}~\bibnamefont {Pothier}}, \bibinfo {author}
  {\bibfnamefont {D.}~\bibnamefont {Esteve}}, \ and\ \bibinfo {author}
  {\bibfnamefont {N.~O.}\ \bibnamefont {Birge}},\ }\href {\doibase
  10.1103/PhysRevB.68.085413} {\bibfield  {journal} {\bibinfo  {journal} {Phys.
  Rev. B}\ }\textbf {\bibinfo {volume} {68}},\ \bibinfo {pages} {085413}
  (\bibinfo {year} {2003})}\BibitemShut {NoStop}%
\bibitem [{\citenamefont {Liu}\ \emph {et~al.}(2016)\citenamefont {Liu},
  \citenamefont {Akis}, \citenamefont {Ferry}, \citenamefont {Bohra},
  \citenamefont {Somphonsane}, \citenamefont {Ramamoorthy},\ and\ \citenamefont
  {Bird}}]{SLiu2016}%
  \BibitemOpen
  \bibfield  {author} {\bibinfo {author} {\bibfnamefont {B.}~\bibnamefont
  {Liu}}, \bibinfo {author} {\bibfnamefont {R.}~\bibnamefont {Akis}}, \bibinfo
  {author} {\bibfnamefont {D.~K.}\ \bibnamefont {Ferry}}, \bibinfo {author}
  {\bibfnamefont {G.}~\bibnamefont {Bohra}}, \bibinfo {author} {\bibfnamefont
  {R.}~\bibnamefont {Somphonsane}}, \bibinfo {author} {\bibfnamefont
  {H.}~\bibnamefont {Ramamoorthy}}, \ and\ \bibinfo {author} {\bibfnamefont
  {J.~P.}\ \bibnamefont {Bird}},\ }\href
  {http://stacks.iop.org/0953-8984/28/i=13/a=135302} {\bibfield  {journal}
  {\bibinfo  {journal} {Journal of Physics: Condensed Matter}\ }\textbf
  {\bibinfo {volume} {28}},\ \bibinfo {pages} {135302} (\bibinfo {year}
  {2016})}\BibitemShut {NoStop}%
\bibitem [{\citenamefont {Ketzmerick}(1996)}]{SPhysRevB.54.10841}%
  \BibitemOpen
  \bibfield  {author} {\bibinfo {author} {\bibfnamefont {R.}~\bibnamefont
  {Ketzmerick}},\ }\href {\doibase 10.1103/PhysRevB.54.10841} {\bibfield
  {journal} {\bibinfo  {journal} {Phys. Rev. B}\ }\textbf {\bibinfo {volume}
  {54}},\ \bibinfo {pages} {10841} (\bibinfo {year} {1996})}\BibitemShut
  {NoStop}%
\bibitem [{\citenamefont {Kantelhardt}(2012)}]{Skantelhardt2012fractal}%
  \BibitemOpen
  \bibfield  {author} {\bibinfo {author} {\bibfnamefont {J.~W.}\ \bibnamefont
  {Kantelhardt}},\ }in\ \href@noop {} {\emph {\bibinfo {booktitle} {Mathematics
  of complexity and dynamical systems}}}\ (\bibinfo  {publisher} {Springer},\
  \bibinfo {year} {2012})\ pp.\ \bibinfo {pages} {463--487}\BibitemShut
  {NoStop}%
\bibitem [{\citenamefont {Grassberger}\ and\ \citenamefont
  {Procaccia}(1983)}]{SPhysRevLett.50.346}%
  \BibitemOpen
  \bibfield  {author} {\bibinfo {author} {\bibfnamefont {P.}~\bibnamefont
  {Grassberger}}\ and\ \bibinfo {author} {\bibfnamefont {I.}~\bibnamefont
  {Procaccia}},\ }\href {\doibase 10.1103/PhysRevLett.50.346} {\bibfield
  {journal} {\bibinfo  {journal} {Phys. Rev. Lett.}\ }\textbf {\bibinfo
  {volume} {50}},\ \bibinfo {pages} {346} (\bibinfo {year} {1983})}\BibitemShut
  {NoStop}%
\bibitem [{\citenamefont {Kellay}\ and\ \citenamefont
  {Goldburg}(2002)}]{SKellay2002}%
  \BibitemOpen
  \bibfield  {author} {\bibinfo {author} {\bibfnamefont {H.}~\bibnamefont
  {Kellay}}\ and\ \bibinfo {author} {\bibfnamefont {W.~I.}\ \bibnamefont
  {Goldburg}},\ }\href {http://stacks.iop.org/0034-4885/65/i=5/a=204}
  {\bibfield  {journal} {\bibinfo  {journal} {Reports on Progress in Physics}\
  }\textbf {\bibinfo {volume} {65}},\ \bibinfo {pages} {845} (\bibinfo {year}
  {2002})}\BibitemShut {NoStop}%
\bibitem [{\citenamefont {Boffetta}\ and\ \citenamefont
  {Ecke}(2012)}]{SBoffetta2012}%
  \BibitemOpen
  \bibfield  {author} {\bibinfo {author} {\bibfnamefont {G.}~\bibnamefont
  {Boffetta}}\ and\ \bibinfo {author} {\bibfnamefont {R.~E.}\ \bibnamefont
  {Ecke}},\ }\href {\doibase 10.1146/annurev-fluid-120710-101240} {\bibfield
  {journal} {\bibinfo  {journal} {Annual Review of Fluid Mechanics}\ }\textbf
  {\bibinfo {volume} {44}},\ \bibinfo {pages} {427} (\bibinfo {year} {2012})},\
  \Eprint
  {http://arxiv.org/abs/https://doi.org/10.1146/annurev-fluid-120710-101240}
  {https://doi.org/10.1146/annurev-fluid-120710-101240} \BibitemShut {NoStop}%
\end{thebibliography}
%

\clearpage

\setcounter{page}{1}

\renewcommand{\figurename}{}
\renewcommand{\thefigure}{Supplementary~Figure~\arabic{figure}}
\setcounter{figure}{0}

\renewcommand{\theequation}{\arabic{equation}}
\setcounter{equation}{0}

\renewcommand{\tablename}{}
\renewcommand{\thetable}{Supplementary~Table~\arabic{table}}
\setcounter{table}{0}

\renewcommand{\thesection}{SUPPLEMENTARY NOTE~\arabic{section}}
\setcounter{section}{0}

{\Large{\textbf{Supplementary Information}  }}

\renewcommand{\figurename}{}
\renewcommand{\thefigure}{Supplementary~Figure~\arabic{figure}}
\setcounter{figure}{0}

\renewcommand{\theequation}{S\arabic{equation}}
\setcounter{equation}{0}

\renewcommand{\tablename}{}
\renewcommand{\thetable}{Supplementary~Table~\arabic{table}}
\setcounter{table}{0}

\renewcommand{\thesection}{SUPPLEMENTARY NOTE~\arabic{section}}
\setcounter{section}{0}

\section{UCF data}

\begin {figure*}[!h]
  \begin{center}
    \includegraphics[width=0.85\textwidth]{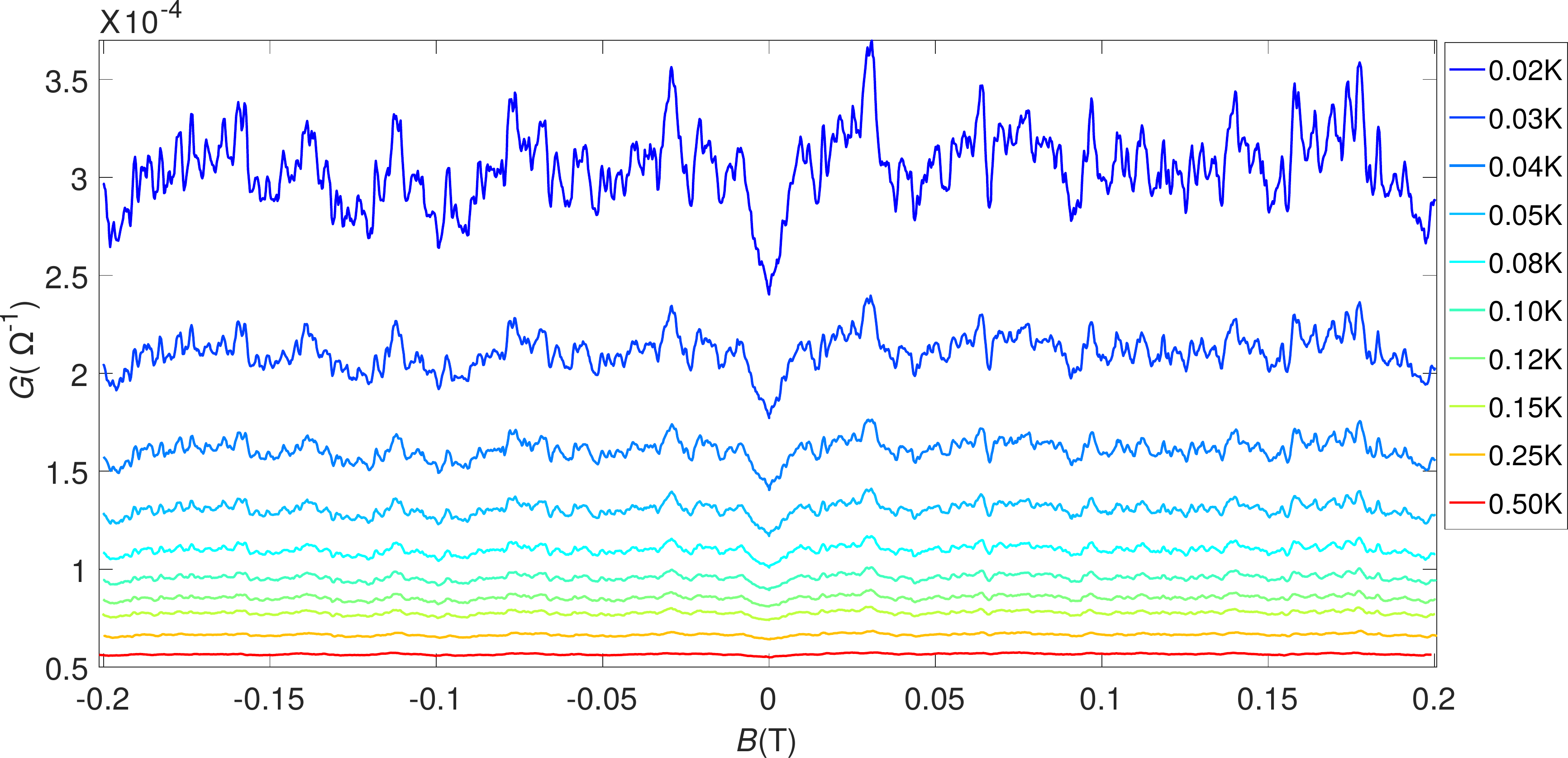}
    \small { \caption{  \textbf{Plot of UCF.} Illustrative plots of the magnetoconductance ($G$) versus magnetic field ($B$), for our device G30M4,  measured at different temperatures, with $\Delta V_{\mathrm{G}} = -0.7$~V. The data have been shifted vertically for clarity.
        \label{fig:df_ucfplot} }}
  \end{center}
\end {figure*}

\ref{fig:df_ucfplot} shows plots of the magnetoconductance at different temperatures ($T$), for our device G30M4. The data have been vertically shifted for clarity. The measurements were carried out at a fixed gate voltage $\Delta V_{\mathrm{G}} = -0.7$~V. 
Each of the magnetoconductance plots show aperiodic oscillations, which are  symmetric in magnetic field ($B$).  
The amplitude of the oscillations gets suppressed with increasing $T$, however, all 
the oscillation features (peaks and dips in the  magnetoconductance) appear at exact same value of  $B$. 
These symmetric and reproducible oscillations are the signatures of UCF.

We  first symmetrize each of the the magnetoconductance 
\begin{equation}
G_{\rm sym}(B) = \frac{G(B) + G(-B)}{2},
\end{equation} 
to remove the anti-symmetric noise contribution from our data. This symmetric component, 
henceforth referred to as $G(B)$, is used for all our subsequent  analysis.


\section{Phase Coherence Length}

 We estimate the phase-coherence length ($L_\upphi$) by  two standard methods, other than obtaining from the UCF variance~\cite{SBeenakker19911},  which provid, qualitatively,  similar result [$L_\upphi$ dependence of $D_{\mathrm{F}}$, shown in Fig. 3(b) of the manuscript].
First, we extract $L_\upphi$ by fitting the weak-localization peak with 
 the Hikami-Larkin-Nagaoka equation~\cite{SPhysRevLett.100.056802},
\begin{equation}
\Delta \sigma = \frac{e^2}{\pi h} \times \bigg[ F \bigg(\frac{8 \pi B}{\Phi_0 L_\upphi^{-2}} \bigg) - F \bigg( \frac{8 \pi B}{\Phi_0 \big[L_\upphi^{-2} + 2L_{\mathrm{i}}^{-2} \big] } \bigg) -2F \bigg( \frac{8 \pi B}{\Phi_0 \big[ L_\upphi^{-2} + L_{\mathrm{i}}^{-2} + L_\ast^{-2} \big] } \bigg)  \bigg]
\label{eqn:wl}
\end{equation}
where $F(z) = \ln z + \psi(0.5 + z^{-1})$, and $\psi(x)$ is the digamma function, 
 $L_{\mathrm{i}}$ and $L_\ast$ are the characteristic length scale for inter-valley scattering and intra-valley scattering, respectively.  The first term in Eqn.~\ref{eqn:wl} gives the weak-localization correction, and the other terms are attributed  to weak anti-localization.

 \begin {figure*}[!ht]
   \begin{center}
     \includegraphics[width=0.55\textwidth]{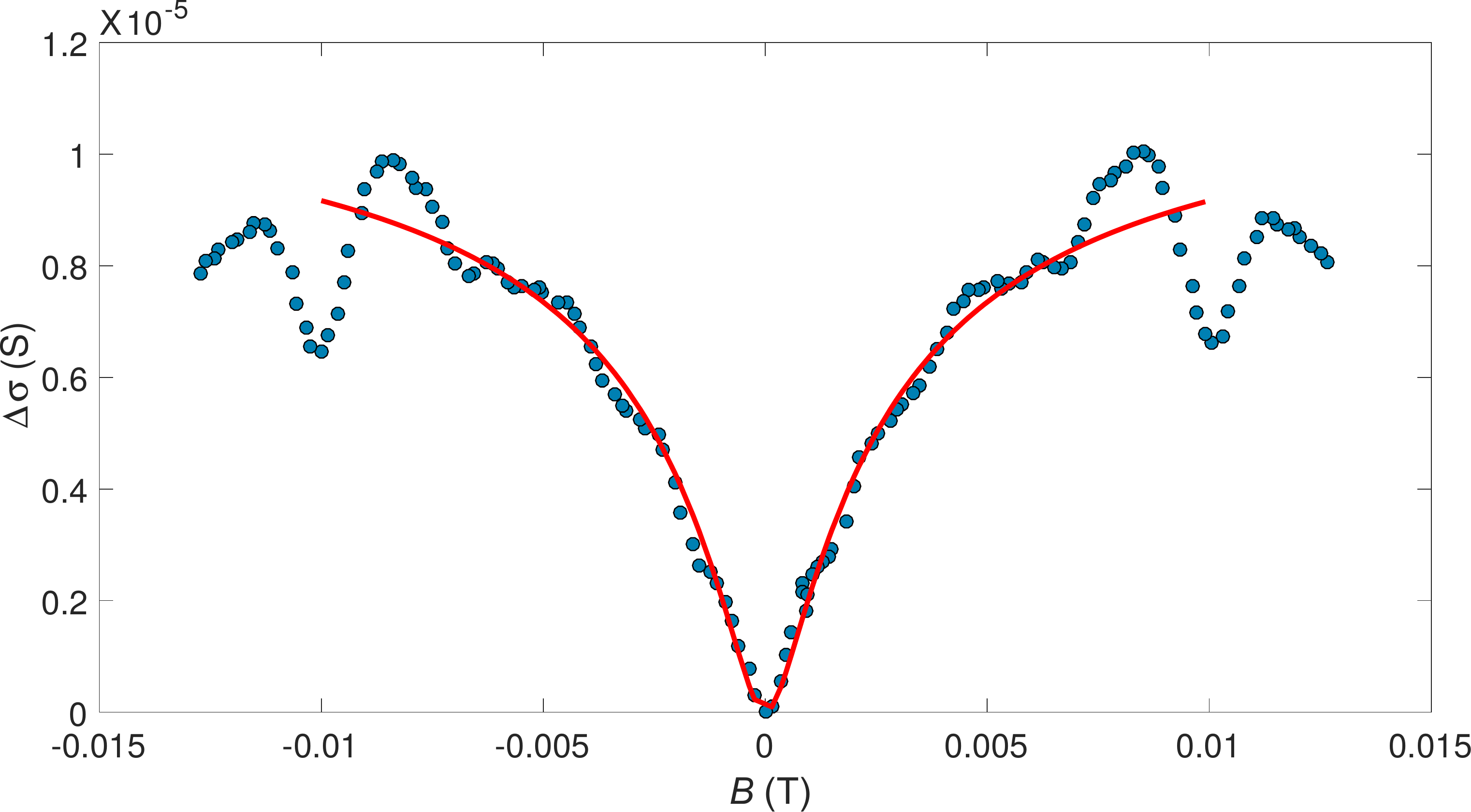}
     \small { \caption{  \textbf{Fit with the HLN equation to extract $L_\upphi$.} A representative plot of  $\Delta\sigma$  versus  magnetic field $B$ is shown here. The filled circles represent the data points and the red thick line is the fit to the data using Eqn.~\ref{eqn:wl}.  \label{fig:wlfit} }}
   \end{center}
 \end {figure*}
\begin {figure*}[!t]
  \begin{center}
    \includegraphics[width=0.525\textwidth]{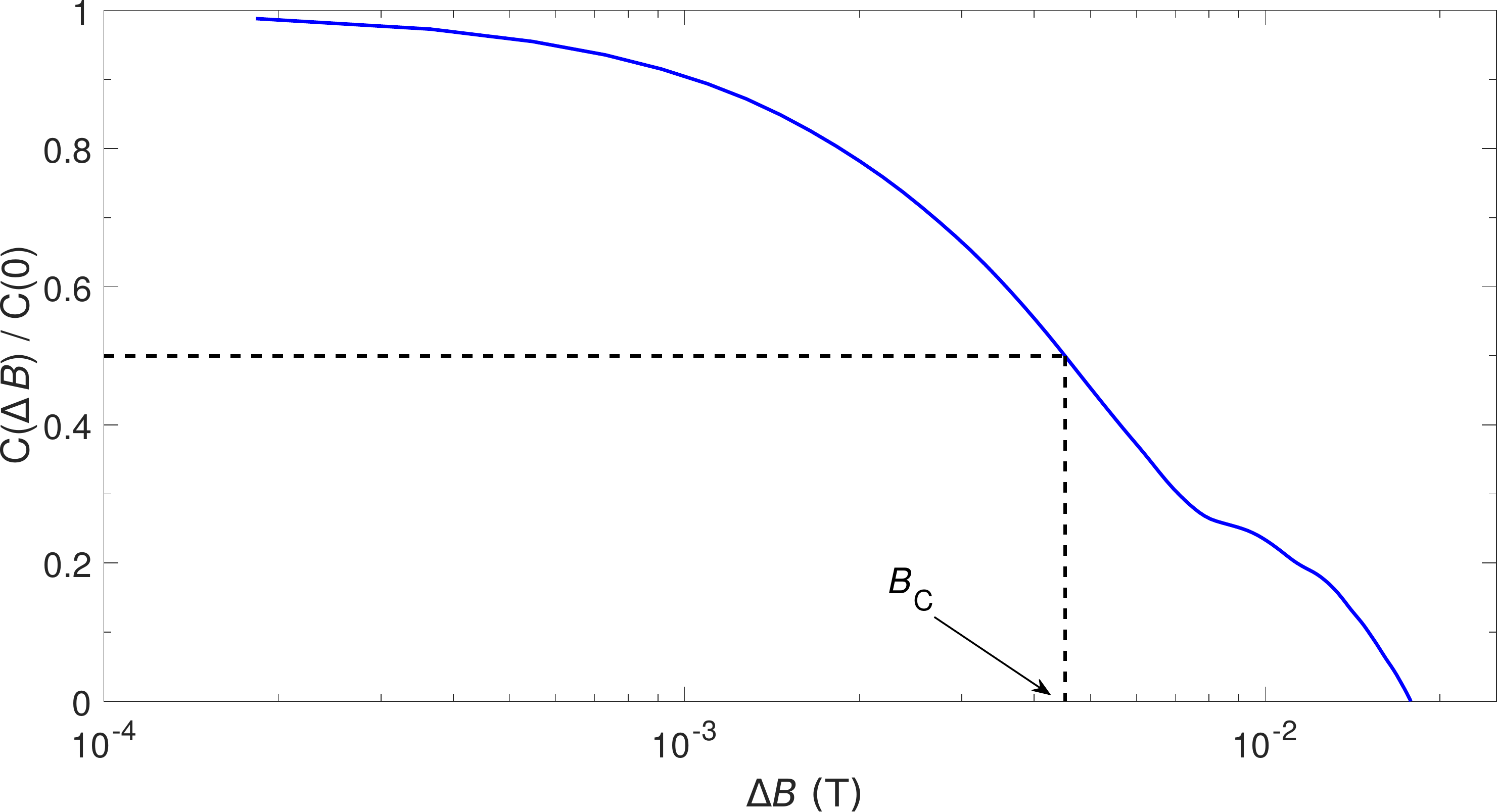}
    \small { \caption{  \textbf{$L_\upphi$ from autocorrelation of UCF.} Plot of normalized autocorrelation  $C(\Delta B)/C(0)$  versus $\Delta B$. The correlation  field  $B_{\mathrm{C}}$, defined as the magnetic field at which  the normalized autocorrelation becomes 0.5, is marked with an arrow. \label{fig:autocorr} }}
  \end{center}
\end {figure*}

\begin{table}[!b]
\caption{ Comparison of $L_\upphi$ extracted via different methods\label{tbl:lphi}}
\begin{tabular}{|c|c|c|c|}
\hline
\textbf{Method} & UCF variance & HLN fit & autocorrelation of UCF  \\
\hline
$L_\upphi$($\mu$m) & 1.20  & 1.27 & 0.96 \\
\hline
\end{tabular}
\end{table}

\ref{fig:wlfit} shows plots of a typical $\Delta\sigma=\sigma(B)-\sigma(0)$ versus $B$, and the fit to the data with Eqn.~\ref{eqn:wl}. The extracted value of  $L_\upphi$   from the fit is $1.27~\mu$m (\ref{tbl:lphi}).

  $L_\upphi$ was also  estimated from the auto-correlation ($C(\Delta B)$) of the UCF~\cite{SBeenakker19911,SHorsell20091041}. 
The autocorrelation of the magnetoconductance is defined as
\begin{equation}
C(\Delta B) = \int dB G(B) G(B+\Delta B) .
\end{equation}

The correlation  magnetic field $B_{\mathrm{C}}$ is then obtained from the autocorrelation $C(\Delta B)$ using the relation,
\begin{equation}
C(B_{\mathrm{C}}) = \frac{1}{2} C(0).
\end{equation}
 $L_\upphi$, is then calculated from $B_{\mathrm{C}}$  using the relation 
\begin{equation}
L_\upphi = \sqrt{h/eB_{\mathrm{C}}}.
\end{equation}

\ref{fig:autocorr} shows plot of normalized autocorrelation $C(\Delta B)/C(0)$ versus $\Delta B$ for the same UCF data [for which the fit with the HLN equation is shown in \ref{fig:wlfit}].  

A comparison of values of $L_\upphi$ extracted from a UCF data, measured at $T=0.2$~mK, using the above-mentioned different methods are shown in \ref{tbl:lphi}.

\section{Effect of Temperature on UCF}

\begin {figure*}[!t]
  \begin{center}
    \includegraphics[width=0.75\textwidth]{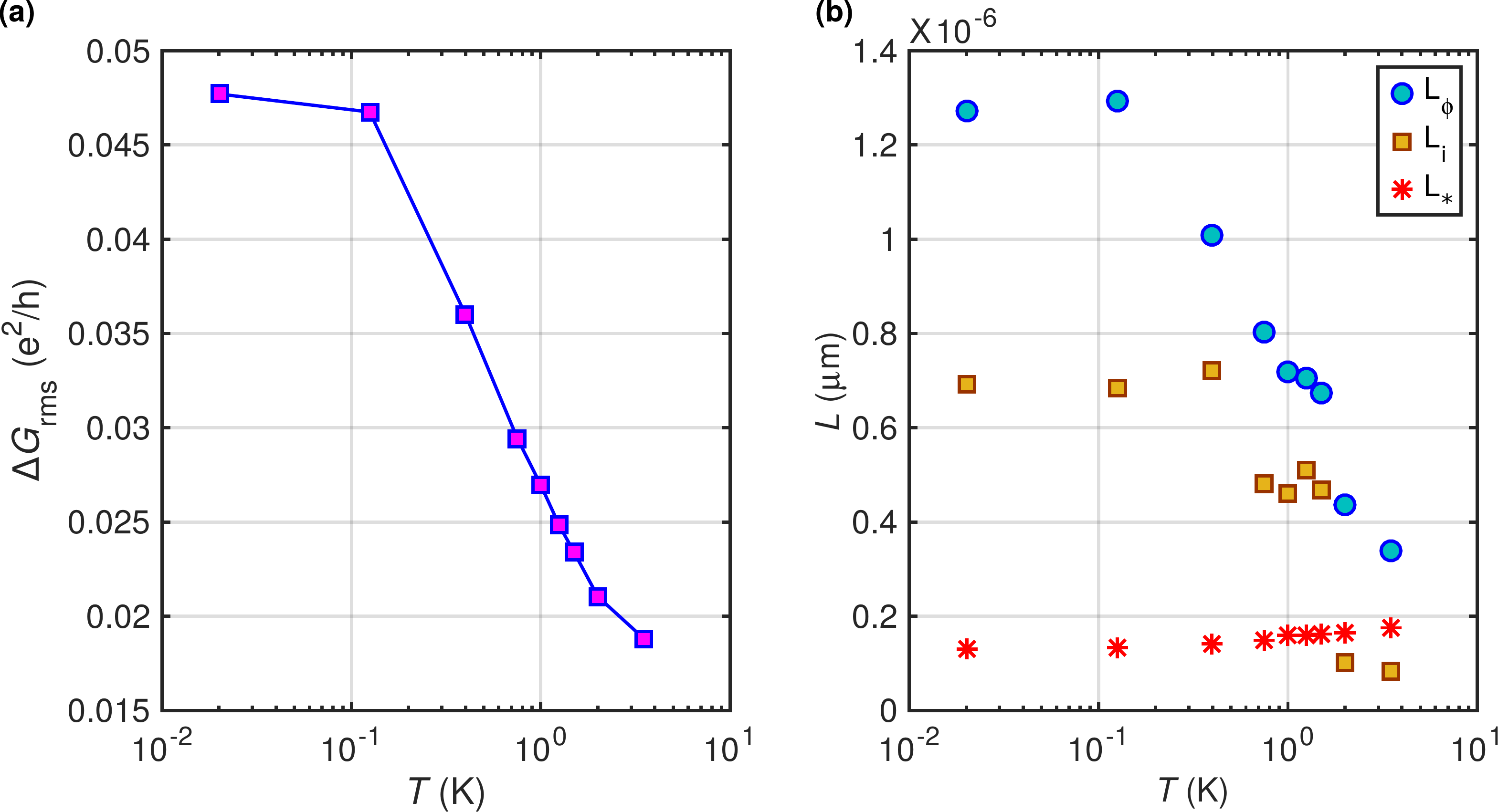}
    \small { \caption{  \textbf{Effect of $T$ on UCF.} (a) Plot of rms value of UCF, measured at $\Delta V_G = $0.2 V, as a function of T.  
        (b) Plot of $L_\upphi$ , $L_{\mathrm{i}}$ and $L_\ast$ extracted from fitting the magnetoresistance data with the HLN equation (Eqn~\ref{eqn:wl}), at $T=$20 mK and $\Delta V_{\mathrm{G}}=$ 0.2 V. 
        \label{fig:lphi_vg} }}
  \end{center}
\end {figure*}

Plot of  the rms value of the magnetoconductance fluctuation ($\delta G_{\rm rms}$) versus $T$  is shown in ~\ref{fig:lphi_vg}(a). The  UCF corresponding to this data were measured at $\Delta V_{\mathrm{G}} = $0.2 V [UCF series shown in Fig. 2(a) of main text]. We observe a strong suppression of the magnetoconductance fluctuations with increasing temperature which is directly attributed to charge carrier dephasing induced by thermal scattering~\cite{S0953-8984-22-20-205301, SBerger1191, SBohra2012, SPhysRevLett.100.056802, SHorsell20091041}.

\ref{fig:lphi_vg}(b) shows plots of $L_\upphi$ (in blue filled circles), $L_{\mathrm{i}}$ (in yellow filled squares) and $L_\ast$ (in red asterisk) versus $T$, extracted from fitting the weak localization correction in the magnetoresistance data with the HLN equation (Eqn.~\ref{eqn:wl}), at $T=20$~mK and $\Delta V_{\mathrm{G}}=0.2$~V. We find that the goodness of the fit is strongly influenced by the presence of UCF, as was observed by other groups as well~\cite{SPhysRevLett.98.176805, SPhysRevLett.100.056802}.

We observe some evidence of a labelling off of  $L_\upphi(T)$ around 100 mK. This is an issue that has been at the forefront of research in several other semiconducting materials including doped Si~\cite{SPhysRevLett.91.066604,SPhysRevB.55.R13452,SPhysRevB.68.085413}. Saturation of $L_\upphi(T)$ have been observed in graphene at a much higher temperatures ($\sim 4$ K) and had been argued to arise from finite-size effects~\cite{SPhysRevLett.100.056802}.
The presence of disorder also enhances the saturation of $L_\upphi$. This is because the presence of disorder leads to the formation of electron-hole puddles, which modify the conducting paths and decrease the effective size of the device~\cite{PhysRevLett.100.056802}. This, in turn, results in a size-limited saturation of $L_\upphi$. Earlier experiments have also shown that, scattering from extremely low concentration of magnetic impurity, even at a level undetectable
by other means, can lead to the widely observed saturation of $L_\upphi$ (or equivalently, the dephasing time $\tau_\upphi$ ) in weakly disordered systems~\cite{SPhysRevB.68.085413}.
Our device has a higher mobility and a longer channel length than  in Ref~\cite{SPhysRevLett.100.056802}.
  Thus we expect the effects of disorder and small channel length, which typically leads to a saturation of $L_\upphi$, to be much weaker in our device. Hence,  it is reasonable to observe a stronger temperature dependence of $L_\upphi$ down to lower temperatures, and a probable saturation only below 0.1~K.

\section{ UCF at different magnetic fields}

We have also measured UCF at a fixed magnetic field, by varying gate-voltage. 
We compare the rms value of the magnetoconductance fluctuations ($\delta G_{\rm rms}$) measured in two different ways: (1) at a constant magnetic field by varying gate voltage, and (2) at a constant gate voltage (with $\Delta V_{\mathrm{G}} \sim$ 0) by varying the magnetic field. We  observe that the magnitude of the UCF measured during the gate sweep is $\sim$ 1.6 times larger than the UCF measured during the field sweep, which indicates non-ergodicity of UCF~\cite{SLiu2016}.

\begin {figure*}[!t]
  \begin{center}
    \includegraphics[width=0.75\textwidth]{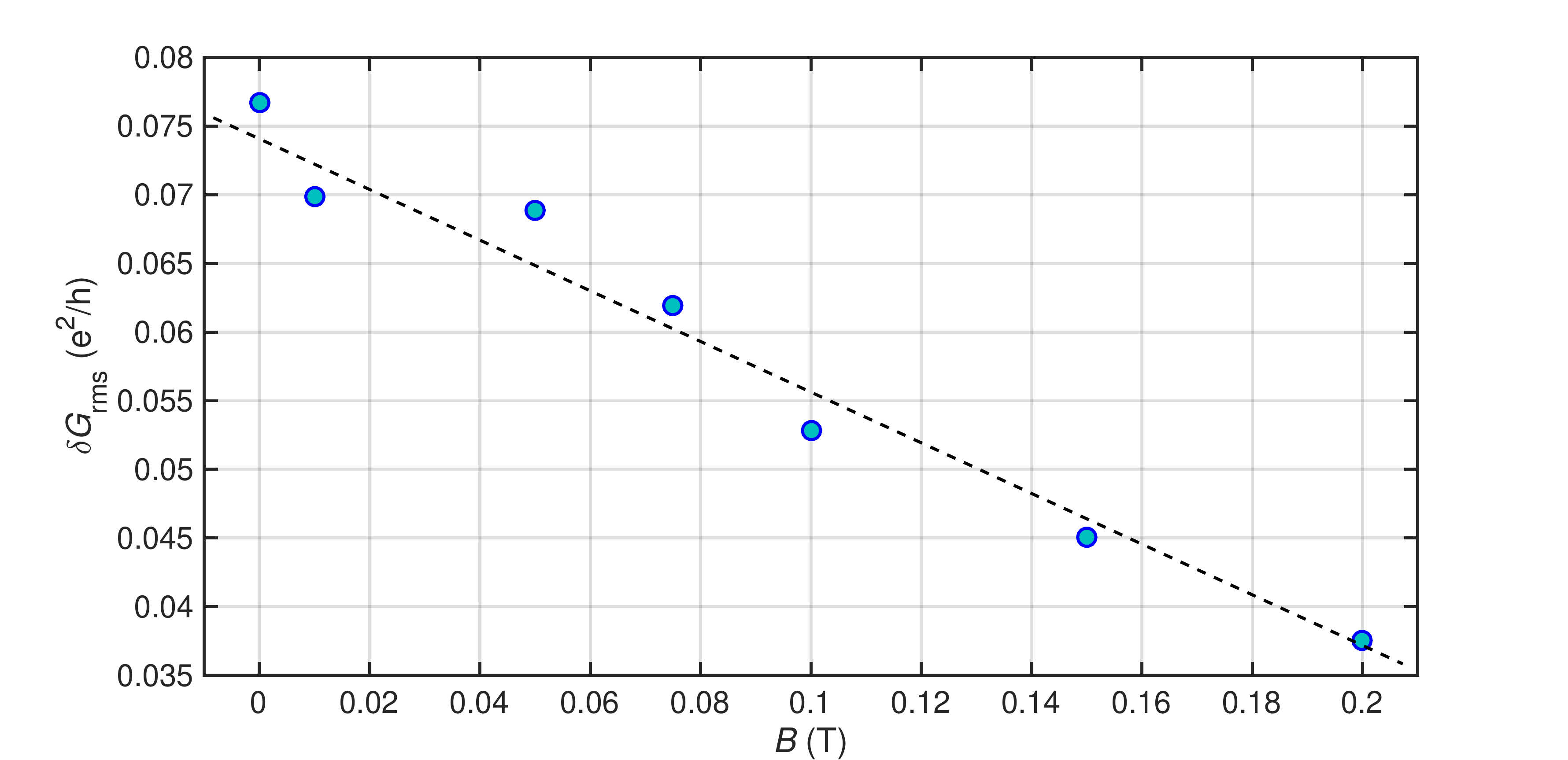}
    \small { \caption{  \textbf{ UCF amplitude versus B.} Typical plot of  $\delta G_{\rm rms}$ versus $B$  - the data were acquired by sweeping the gate voltage at a fixed magnetic field. The data were acquired at $T=$20 mK. Blue filled circles represent the data and the dashed line is guide to the eye. 
        \label{fig:rms_b} }}
  \end{center}
\end {figure*}

Furthermore,  we have carried out UCF measurements by sweeping the gate voltage at different values of the magnetic field in the range of 0 T to 0.2 T. In \ref{fig:rms_b}, we plot the rms value of conductance fluctuations,  $\delta G_{\rm rms}$ {versus $B$}. We find that $\delta G_{\rm rms}$ decreases with increasing $B$. These observations are  in excellent agreement with predictions of behaviour of UCF in the presence of long-range component of random scattering potential~\cite{SLiu2016}.

\section{Calculation of  fractal dimensions}

\begin {figure*}[!h]
  \begin{center}
    \includegraphics[width=0.7\textwidth]{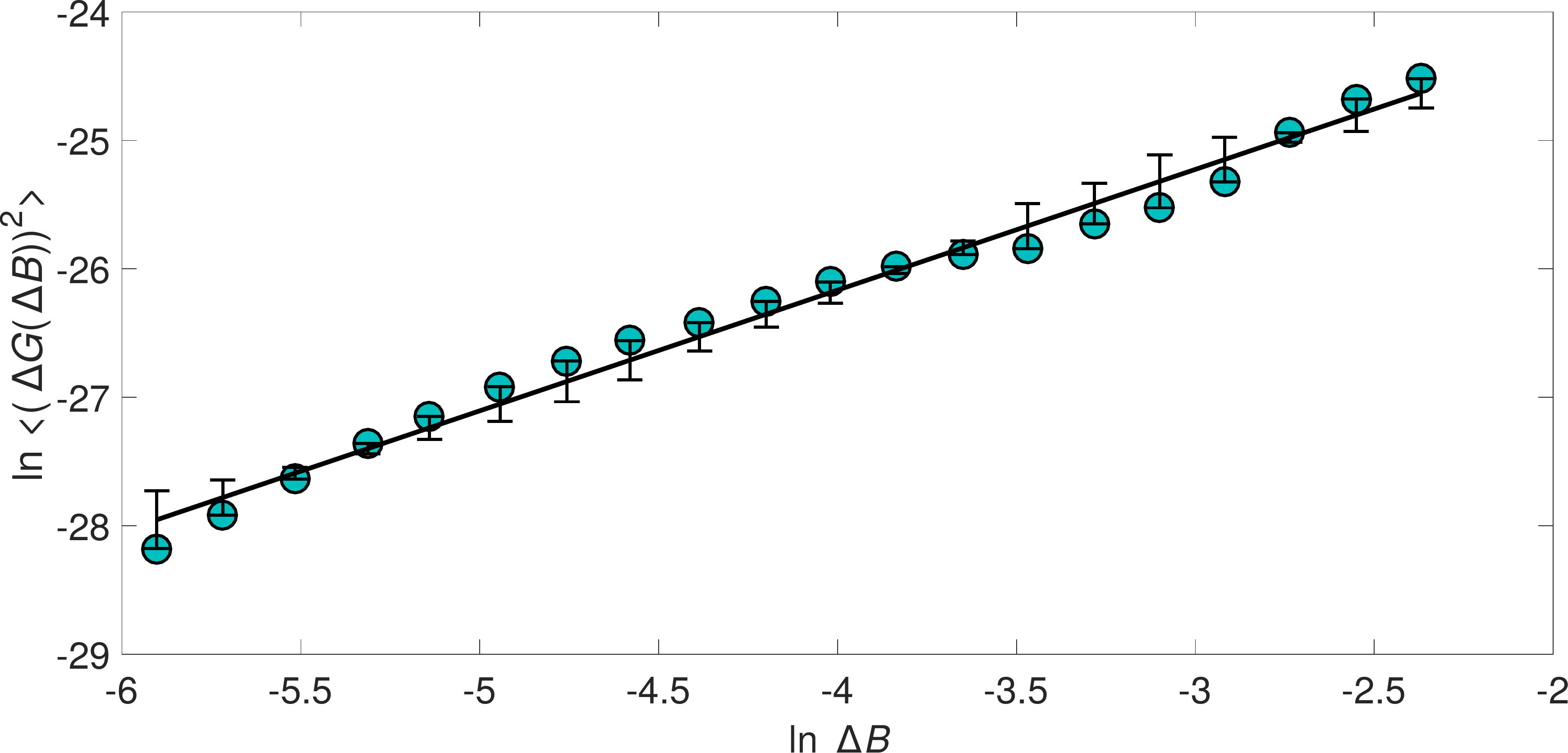}
    \small { \caption{\textbf{Fractal scaling plot.} A representative, ln-ln plot of $\langle (\Delta G(\Delta B))^2 \rangle$ versus $\Delta B$ from a typical UCF. 
        The filled circles are the data from our experiment and the thick black line connecting them is the linear best-fit that yields $D_{\mathrm{F}} \simeq 1.5$. The error-bars represent deviation of the data-points from the best-fit line.
        \label{fig:df_scaling} }}
  \end{center}
\end {figure*}

We calculate the fractal dimensions, from our UCF data, in the following way. 

We begin by defining and calculating $\delta G(B)$ = $G(B) - \langle G(B) \rangle$, where the angular brackets define the average of over $B$.  To calculate the fractal dimension, we begin~\cite{SPhysRevB.54.10841} by dividing our UCF data into $N_s$ overlapping segments of size $\Delta B$. We then define $\Delta G(\Delta B) \equiv \delta G(B) - \delta G(B+\Delta B)$ from which we calculate $\langle (\Delta G(\Delta B))^2 \rangle $, where we average over all the  $N_s$ segments.  Thence we evaluate $\langle (\Delta G(\Delta B))^2 \rangle $ for different values of  $\Delta B$, and study the scaling.  Typically, $\langle (\Delta G(\Delta B))^2 \rangle ~\sim ~(\Delta B)^\gamma$. From fits to log-log plots of  $\langle (\Delta G(\Delta B))^2 \rangle$ versus $\Delta B$  (see \ref{fig:df_scaling} for  a representative example),  we obtain from our data the value of $\gamma$, which yields the fractal dimension $D_{\mathrm{F}}$ via $D_{\mathrm{F}} = 2 - \gamma /2$; for the data shown in \ref{fig:df_scaling}, this yields $D_{\mathrm{F}} \simeq 1.5$.

\section{Calculating multifractal exponents}

\begin {figure*}[!t]
  \begin{center}
    \includegraphics[width=1.0\textwidth]{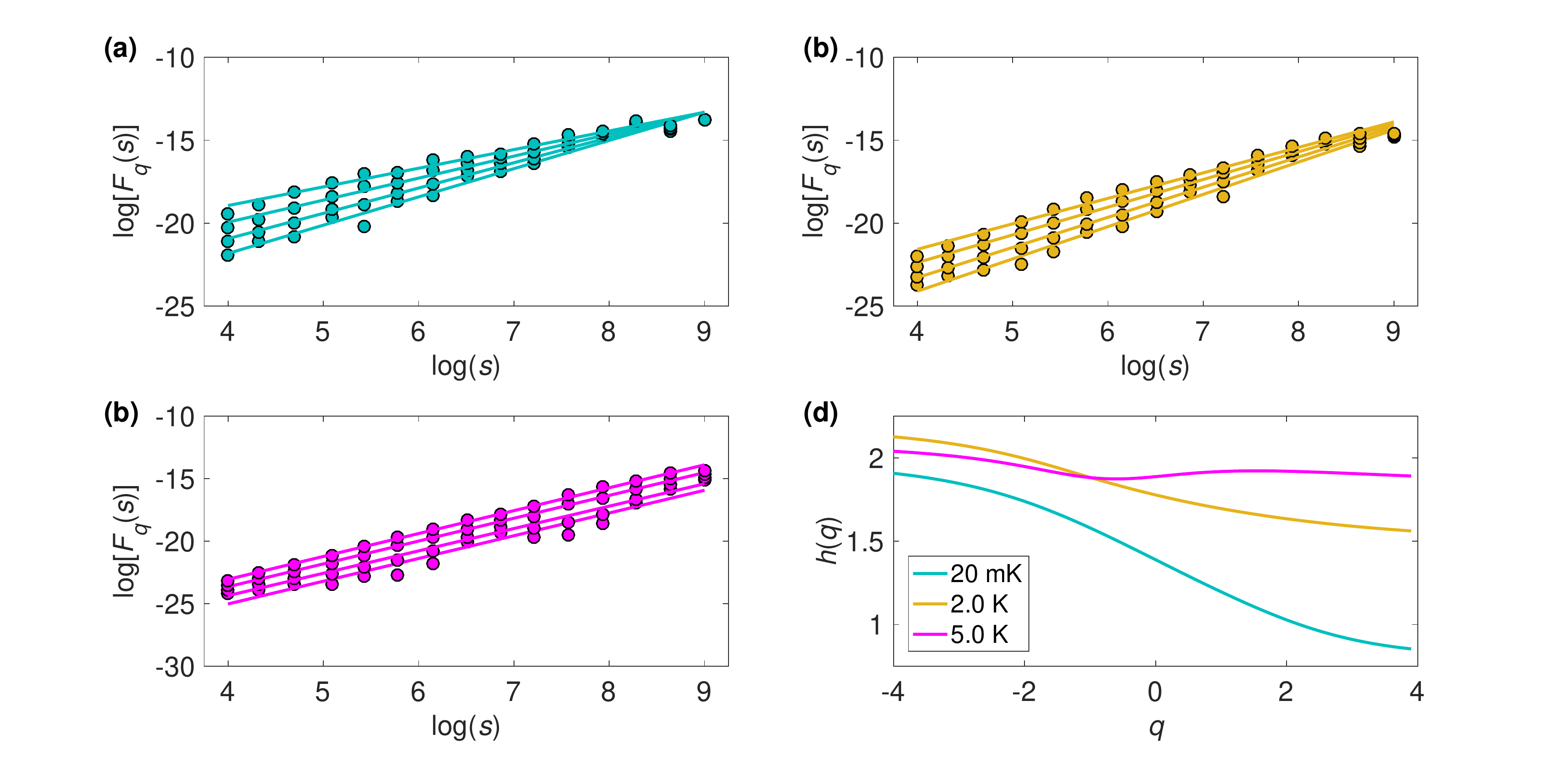}
    \small { \caption{\textbf{Multifractal scaling plot.} (a)-(c) Plots of $\log[F_q(s)]$ [Eqn. \ref{eqn:fqs}] as a function of scale size $\log(s)$ for different orders $q$. The text inside the figures indicate the temperature at which the 
        measurements were made. (d) A plot of the order-$q$ scaling exponent $h(q)$ as a function of $q$; these exponents have  been  extracted from the plots (a)-(c), 
        as discussed in the text. \label{fig:fq} }}
  \end{center}
\end {figure*}

We compute the multifractal singularity spectrum form our UCF by using the method of Multifractal Detrended Fluctuation Analysis (MFDFA)~\cite{Skantelhardt2012fractal}, which we describe below. 

We begin by a {\it local detrending} of our data. To do this, we choose a segment  of the magnetic field $\Delta B$ and then divide the UCF into different overlapping segments.  We then fit the different segments of our conductance data ${G}$ with a polynomial $P_m$ and remove the 
local trend:  
\begin{equation}
g(i) = G(i) - P_m(i). 
\end{equation}
It is important to keep the same order $m$ of the polynomial throughout this analysis and for all the segments. Alternatively, 
we can divide the conductance data  ${G}$, with  $N$ data points, into $N_s$ segments: $N_s = int(N/s)$, with $s$ points in each 
segment.  In our analysis, the detrending is done with a polynomial of order 1.

Next, we evaluate the root-mean-square (rms) fluctuations and the order-$q$  moment of these fluctuation via 
\begin{equation}
g_{\rm rms} = \left (\frac{1}{s} \sum_{i=1}^s g(i)^2\right )^{1/2}. 
\end{equation}

We then define
\begin{equation}
F_q(s) = \left (\frac{1}{N_s} \sum_{j=i}^{N_s} g_{\rm rms}(j)^q\right )^{1/q}. 
\label{eqn:fqs}
\end{equation}
In \ref{fig:fq}(a-c) we show representative (log-log) plots of $F_q$ versus $s$ for three different temperatures and for different values of $q$. From such plots it is possible to calculate the  generalized order-$q$  scaling exponent through the scaling relation 
\begin{equation}
F_q(s) \sim s^{H(q)}. 
\label{eqn:Hq}
\end{equation}
We calculate the order-$q$  scaling exponent $H(q)$ for different values of $q$ and at different temperatures. In \ref{fig:fq}(d) we show a representative plot of $H(q)$ versus $q$ from data at the  three temperatures shown in \ref{fig:fq}(a-c).

\begin {figure*}[!t]
  \begin{center}
    \includegraphics[width=.75\textwidth]{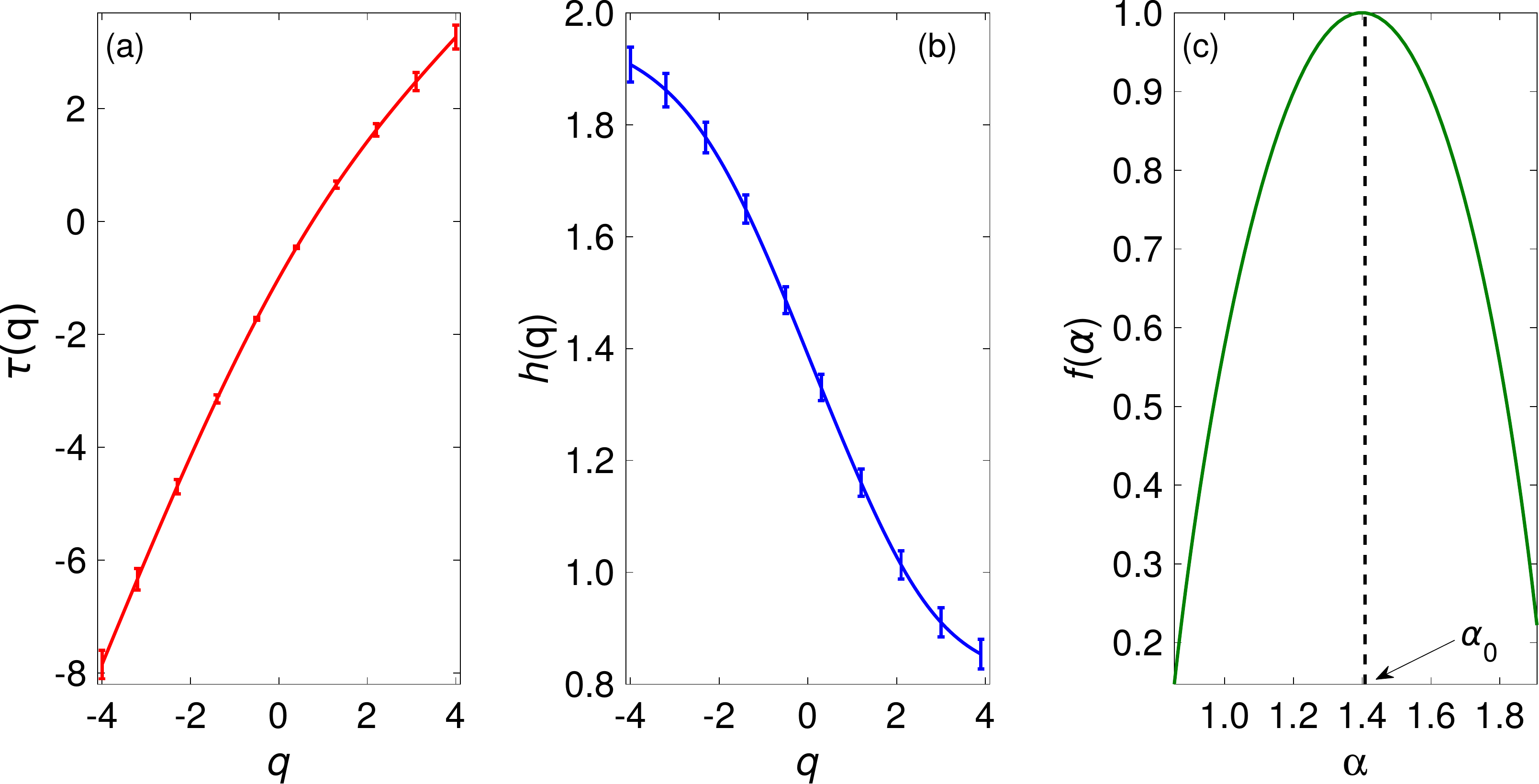}
    \small { \caption{\textbf{Multifractal spectrum.} Plots of (a)  $\tau(q)$ versus $q$ and (b)  $h(q)$ versus $q$ for a representative UCF. A few representative error bars shown in the plots represent the standard deviation in $\tau(q)$ and $h(q)$, respectively. (c) The singularity spectrum $f(\alpha)$ versus $\alpha$, obtained from  the Legendre transform of $h(q)$ versus $q$,  is shown. The black dashed vertical line marks the  location of the maximum of $f(\alpha)$.
        \label{fig:mf_exp} }}
  \end{center}
\end {figure*}

We now calculate the singularity spectrum as follows: 
We first calculate 
\begin{equation}
\tau(q) =  H(q) q - 1. 
\end{equation}

From this, we calculate the generalized Hurst exponent
\begin{equation}
h(q) = \frac{d \tau(q)}{ dq}.  
\label{eqn_hq}
\end{equation}
In \ref{fig:mf_exp}(a) we show plot of $\tau (q)$,  for a representative UCF. By using this, and making use of the relation between $\tau(q)$ and  $h(q)$ shown above, we  obtain plots of $h(q)$ such as the one shown in \ref{fig:mf_exp}(b).

$\alpha$ and $f(\alpha)$ follow from $\tau(q)$ and $h(q)$:
\begin{equation}
\alpha = \frac{d \tau(q)}{dq} = h(q);
\end{equation}
\begin{equation} 
f(\alpha) = \alpha q -  \tau(q).
\end{equation}

In \ref{fig:mf_exp}(c) we show a plot of the multifractal singularity spectrum $f(\alpha)$, which shows a clear maximum  at $f(\alpha_0) = 1$. It is important to keep in mind that the position of this maxima $\alpha_0 \equiv h(0)$ depends on both the gate voltage $V_{\mathrm{G}}$ and the temperature $T$. 

The error in $\tau(q)$ was estimated from the fitting error of the plot of $F_q(s)$ versus $s$ [Eqn. \ref{eqn:Hq}]. The error in $h(q)$ was obtained from error in $H(q)$ by using Eqn.~\ref{eqn_hq}. Absolute errors in $\tau(q)$ and $h(q)$ are represented using error-bars in \ref{fig:mf_exp}(a) and (b), respectively. Finally, the error in the spectral width $\Delta \alpha$ is obtained by adding the error in $h(q)$,  obtained at the two end values of q, \textit{i.e.} $q=-4$ and $q=4$.

\section{Multifractality of a randomly shuffled series}

We generate a randomly shuffled sequence following the  prescription mentioned below:

(i) We consider a data sequence $\{ g(i) \}$, $i=1,2,...,N$.

(ii) We then generate a sequence $\{ I(i) \}$ of length $N$, which is a random permutation of the positive integers from $1$ to $N$.

(iii) We then generate a new sequence $\{ g_{\rm rs}(i) \}$, where $g_{\rm rs}(i) = g(I(i))$. 

(iv) We repeat this shuffling for several times, to obtain the final randomly shuffled sequence.

\begin {figure*}[!t]
  \begin{center}
    \includegraphics[width=.75\textwidth]{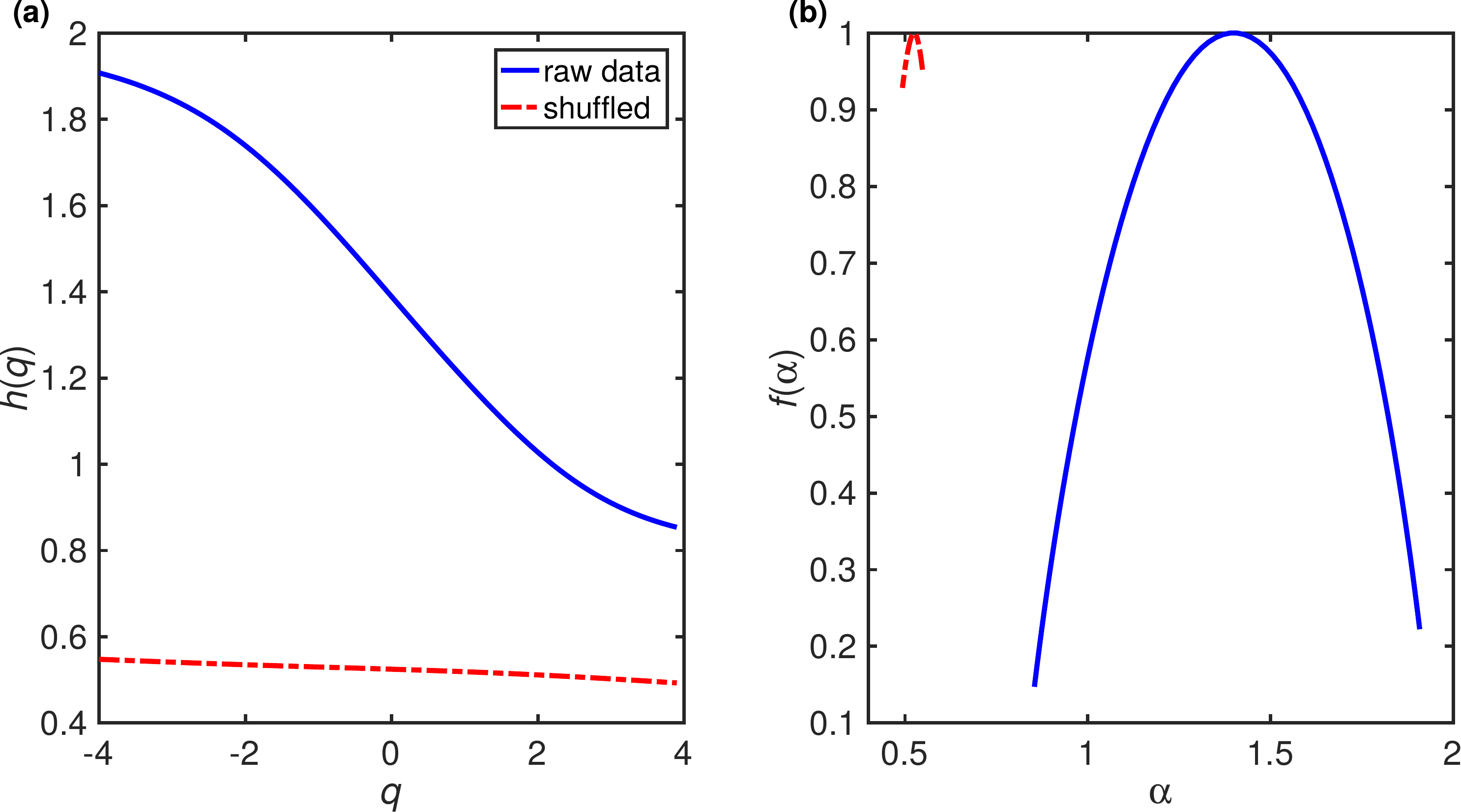}
    \small { \caption{\textbf{Multifractality after shuffling.} Plots of multifractality spectrum (a) $h(q)$ and (b) $f(\alpha)$ of an unshuffled  UCF series (in blue continuous line) and a randomly shuffled  series (in red, dashed line).  
        \label{fig:mf_shuffle} }}
  \end{center}
\end {figure*}

 \ref{fig:mf_shuffle} shows plots  the generalized Hurst exponent  $h(q)$ and the singularity spectrum $f(\alpha)$ of a UCF series (measured at $T$=20~mK and $\Delta V_{\mathrm{G}} = 0.2~$V) and  its randomly shuffled counterpart. The spectrum for the original data series is shown in blue continuous line, where as that for the randomly shuffled series is shown in red dashed line. For the original data series, $h(q)$ has a large range of values and the singularity spectrum $f(\alpha)$ is wide ($\Delta \alpha= 1.05$ ) showing that the data series is multifractal.
On the other hand, for the shuffled series, $h(q)$ is almost $q$-independent and $f(\alpha)$ is extremely narrow ($\Delta \alpha=0.05$), indicating a clear suppression of multifractality.

\section{Monofractality  versus Multifractality}

\begin {figure*}[!h]
  \begin{center}
    \includegraphics[width=0.65\textwidth]{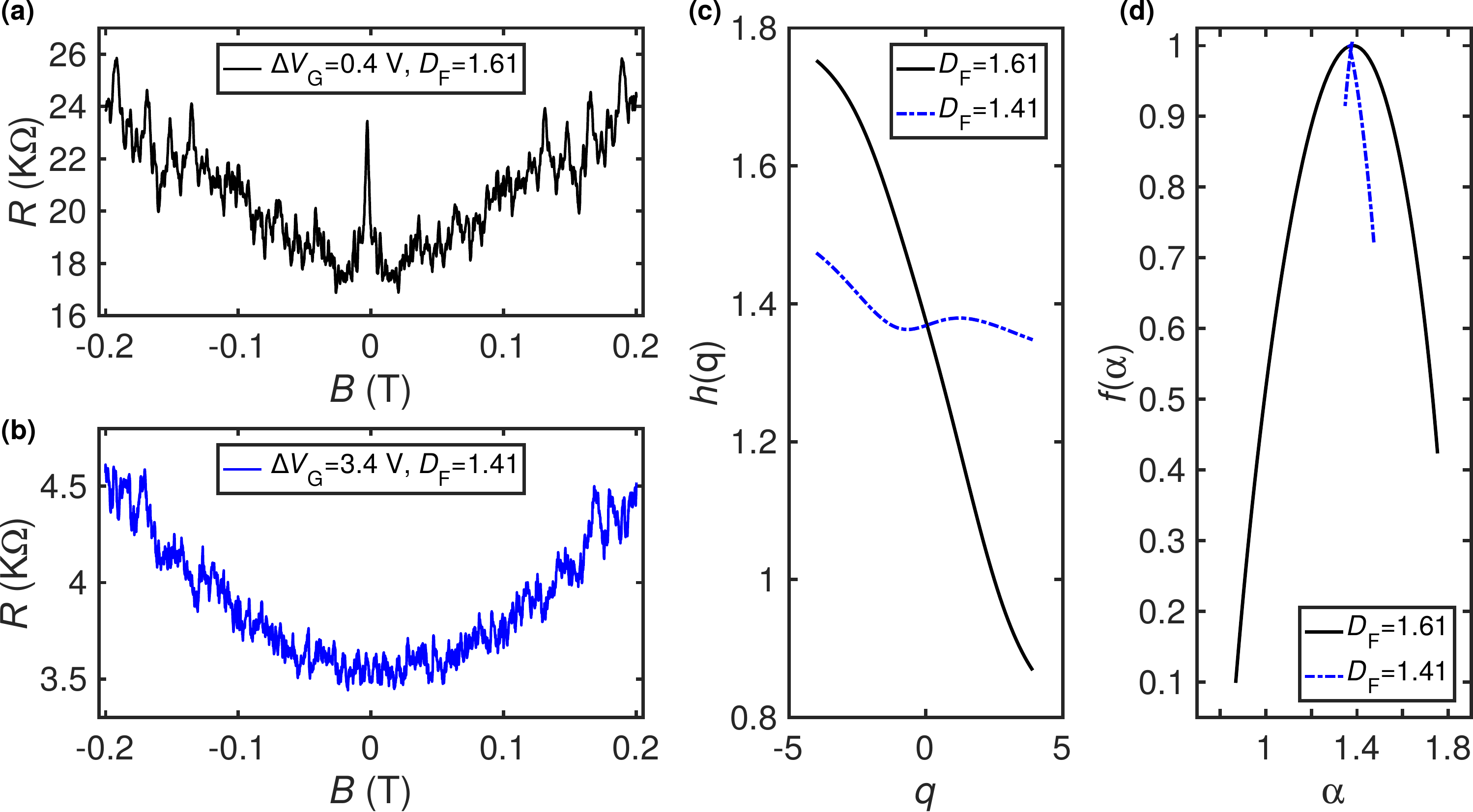}
    \small { \caption{  (a) and (b) show plots of UCF measured on the single-layer graphene device G28M6 at two different values of $\Delta V_{\mathrm{G}}$, 0.4~V and 3.4~V respectively. The data were acquired at $T=20$~mK.  (c) Plots of  the order-$q$ scaling exponent $h(q)$ as a function of $q$; these exponents have  been  extracted from the plots (a) and (b). (d) Plots of the singularity spectrum  $f(\alpha)$ versus $\alpha$, obtained from  the Legendre transform of $h(q)$ versus $q$. Although the UCF measured at both these gate voltages are monofractal,  it is multifractal only for the UCF measured very close to the Dirac point.   
        \label{fig:notallfractalmultifractal2}} }
  \end{center}
\end {figure*}

A n\"aive characterization of the fractal property of a curve, say by the measurement of one fractal dimension, does not rule out multifractality of this curve, which requires the calculation of an infinite number of dimensions~\cite{SPhysRevLett.50.346} (related to our $h(q)$). For monofractal scaling, a single dimension suffices (as, \textit{e.g.}, in the scaling of velocity structure functions in the inverse-cascade region of forced, two-dimensional fluid turbulence~\cite{SKellay2002, SBoffetta2012}).

We  note here, that for the single-layer graphene devices studied by us, even at the lowest temperature, we find $f(\alpha)$ depends on $\Delta V_{\mathrm{G}}$. All the UCFs measured by us are not multifractal; they are multifractal only in a narrow range of gate-voltages around the Dirac point. We show in \ref{fig:notallfractalmultifractal2} an example of  a UCF, that is monofractal, but not multifractal. A UCF measured near the Dirac point ($\Delta V_{\mathrm{G}} =0.4$~V)  [\ref{fig:notallfractalmultifractal2}(a)] is multifractal, as seen in \ref{fig:notallfractalmultifractal2}(c-d) (black solid lines). On the other hand, UCF  measured away from the Dirac point ($\Delta V_{\mathrm{G}} =3.4$ V) [\ref{fig:notallfractalmultifractal2}(b)],    despite having a fractal dimension $D_{\mathrm{F}} = 1.41$, is barely multifractal, as seen from the \textit{almost} $q$ independent $h(q)$ and an extremely narrow singularity spectrum [\ref{fig:notallfractalmultifractal2}(c-d) (blue dashed lines)].

\newpage

\section*{References}
%

\end{document}